# Effects of Influential Points and Sample Size on the Selection and Replicability of Multivariable Fractional Polynomial Models

Willi Sauerbrei[1*], Edwin Kipruto[1*], James Balmford[1+],

[1]Institute of Medical Biometry and Statistics, Faculty of Medicine and Medical Center - University of Freiburg, Germany

*Joint first authorship; + deceased

**Abstract**

The multivariable fractional polynomial (MFP) procedure combines variable selection with a function selection procedure (FSP). For continuous variables, a closed test procedure is used to decide between no effect, linear, FP1 or FP2 functions. Influential observations (IPs) and small sample size can both have an impact on a selected fractional polynomial model. In this paper, we used simulated data with six continuous and four categorical predictors to illustrate approaches which can help to identify IPs with an influence on function selection and the MFP model. Approaches use leave-one or two-out and two related techniques for a multivariable assessment.

In seven subsamples we also investigated the effects of sample size and model replicability. For better illustration, a structured profile was used to provide an overview of all analyses conducted. The results showed that one or more IPs can drive the functions and models selected. In addition, with a small sample size, MFP might not be able to detect non-linear functions and the selected model might differ substantially from the true underlying model. However, if the sample size is sufficient and regression diagnostics are carefully conducted, MFP can be a suitable approach to select variables and functional forms for continuous variables.

# 1. Introduction

In modeling observational data aimed at identifying predictors of an outcome and gaining insight into the relationship between the predictors and the outcome, the process of building a model for description consists of two components: variable selection to identify the subset of "important" predictors, and identification of possible non-linearity in continuous predictors. The ultimate aim is to build a model which is satisfactory in terms of model fit, interpretable from the subject-matter point of view, robust to minor variations in the current data, predictive in new data, and parsimonious (Royston and Sauerbrei 2008).

In model building, many researchers typically assume a linear function for continuous variables (perhaps after applying a "standard" transformation such as log) or divide the variable into several categories. If the assumption of linearity is incorrect, it may prevent the detection of a stronger effect or even cause the effects to be mismodeled. Categorization of continuous variables, which has the effect of modeling (implausible) step functions, is common but widely criticized (Altman et al. 1994; Greenland 1995; Royston et al. 2006; Royston and Sauerbrei 2008) and will not be considered further.

Fractional polynomials have been proposed as a simple method of dealing with non-linearity (Royston and Altman 1994; Sauerbrei and Royston 1999; Sauerbrei et al. 2007; Royston and Sauerbrei, 2008). First-degree (FP1, single power) functions are monotonic, whereas second-degree (FP2, two powers) functions can represent a variety of curve shapes with a single maximum or minimum. Models with degree higher than two are rarely required in practice. Fractional polynomials can be viewed as a compromise between conventional polynomials (e.g., quadratic functions) and nonlinear curves generated by flexible modeling techniques such as spline functions, but without the inflexibility of the former or the potential instability of the latter. FPs are global functions that cannot handle local features, unlike several "flavors" of splines, e.g., restricted regression splines (Harrell 2015), penalized regression splines (Wood 2017), smoothing splines (Hastie and Tibshirani 1990) and p-splines (Eilers and Marx 1996). FPs are relatively stable than local-influence models, which have a higher capacity for model fit but lower transferability and relative instability (Perperoglou et al. 2019; Sauerbrei et al. 2020).

The multivariable fractional polynomial (MFP) approach combines backward elimination with a three step closed test procedure (the function selection procedure, or FSP) to select the most appropriate functional form for continuous variables from the proposed class of fractional polynomial functions (8 FP1 and 36 FP2). In this paper, several issues that may affect the identification and estimation of non-linear functions as well as model replicability was considered. The presence of covariate outliers, or influential points (IPs), may have an undue influence on the chosen model. Diagnostic plots were used to show how to identify influential points. Although influential points can have a strong effect on a selected model (Royston and Sauerbrei 2007), it seems that this issue is widely ignored. We are not aware of any paper discussing the role of influential points in the selection of variables and functional forms for continuous variables.

In addition to the approach in the book by Royston and Sauerbrei (2008), we discussed an extension to considering pairs of influential observations and proposed two approaches for identifying influential points in multivariable models. We concentrated on the identification of IPs and illustrated their effects on functions and models selected by comparing results for data with and without IPs. IPs were eliminated and potential ways (e.g., truncation or preliminary transformation) to handle IPs in real data were not discussed. In real world data, handling of IPs depends strongly on the specific study and main aim of a model. We also considered model replicability across datasets. This is an important aspect of multivariable modeling, particularly in the context of influential points, where the presence of an extreme value of a single covariate may affect the functions selected for that variable, correlated variables, and the overall model. Finally, the effect of sample size was investigated since the selection of variables and functions within the MFP procedure uses test statistics which depend strongly on sample size. In small samples, variables with moderate or weak effects may be incorrectly eliminated or linear functions may be chosen instead of more realistic non-linear functions.

To assess whether MFP selects the "true" underlying model or a model which is close to it, it is imperative to use simulated data in which the parameters are known. In this paper, we used data from the ART study (ART denoting "artificial", Royston and Sauerbrei 2008, Chap.10) which consisted of 5,000 simulated observations. A subset of the ART data ($n$=250) was used as the 'main' dataset to illustrate details of the investigation for influential points. These investigations were also conducted in additional subsets (3 datasets, each of 250 observations) to examine function replicability and the influence of sample size (3 datasets of 125, 250 and 500 observations, respectively) but only selected parts are shown, see 'Data not shown' in Table A1. Based on the key principles of plasmode data sets (Gadbury et al. 2008) the distribution of the predictors in the ART study and their correlation structure was informed by a real study from the German Breast Cancer Study Group (GBSG), as described in a number of earlier publications (Sauerbrei and Royston 1999; Royston and Sauerbrei 2008). More background on the GBSG study, the original data and data of the ART study is available at

http://portal.uni-freiburg.de/imbi/Royston-Sauerbrei-book/index.html#datasets.

To improve the quality of reporting and provide a suitable overview of all analyses conducted, we extended the recently proposed ADEMP structure for simulation studies (Morris et al. 2019) with a structured display of analysis strategies and presentations, named MethProf-simu profile (see Table A1 in the Appendix).

The rest of the paper is organized as follows. Section 2 introduces the MFP approach, while section 3 discusses various aspects of investigations for influential points, model replicability and sample size. Section 4 introduces the simulated data. The results of several investigations for these data are presented in section 5, followed by a discussion and conclusions. Several papers and a book have been published about MFP modelling. Therefore, we provide only a short explanation in the main text and give more details in the Web Appendix A1, intended for readers who are unfamiliar with the approach. Due to space limitations, many analyses and a case study has been relegated to the Web Appendix.

## 2. The Multivariable Fractional Polynomial procedure

MFP is a multivariable model building approach which retains continuous predictors as continuous, finds non-linear functions if they are sufficiently supported by the data, and eliminates predictors with weak or no effects by backward elimination (BE) (Royston and Sauerbrei 2008). The two key components are variable selection with backward elimination and the function selection procedure (FSP) which selects an FP function for each continuous variable. The analyst must decide on a nominal significance level ($\alpha$) for both components. The choice of these two significance levels has a strong influence on the complexity and stability of the final model (Royston and Sauerbrei 2003; Royston and Sauerbrei 2008). The same $\alpha$ level can be used for the two components, though it can differ. This decision strongly depends on the aim of the analysis. In MFP terminology, MFP(0.05) means an MFP model with both variables and functions selected at the 0.05 significance level while MFP(0.05, 0.01) means that variables are selected at the 0.05 level and functions at 0.01 level. In this paper, $\alpha$ = 0.05 was used for both components, but we also showed the threshold values for $\alpha = 0.01$ and in some cases we discussed the result for this significance level in order to illustrate the importance of the chosen significance level on the identification of influential points and on the final model chosen. In principle, the MFP approach prefers simpler models because they transfers better to other settings and are more suited for practical use. This contrasts with local regression modeling (e.g., splines, kernel smoothers etc.) which often starts and ends with more complex models (Sauerbrei et al. 2007).

The class of fractional polynomial (FP) functions is an extension of power transformations of a variable. For most applications, FP1 and FP2 functions are sufficient, and in this paper we allowed FP2 to be the most complex function. For more details, see Royston and Altman (1994), Royston and Sauerbrei (2008) and the MFP website http://mfp.imbi.uni-freiburg.de/.

Fractional Polynomial functions are defined in the following way:

$FP1: \beta x^{p1}$

$FP2: \beta_1 x^{p1} + \beta_2 x^{p2}$,

with exponents $p1$ and $p2$ derived from a set $s = \{-2, -1, -0.5, 0, 0.5, 1, 2, 3\}$, where 0 stands for natural logarithm of x. If $p1 = p2$ (repeated powers), the $FP2$ function is defined as $\beta_1 x^p + \beta_2 x^p \log(x)$. Overall, the set of powers permits 44 models of which 8 are FP1 and 36 are FP2. The FP2 with powers ($p1 = 1, p2 = 2$) is equivalent to the quadratic function. While the permitted class of FP functions appears small, it includes very different types of shapes as illustrated in Figure 1 for the eight FP1 powers and a subset of FP2 powers (Royston and Altman 1994; Royston and Sauerbrei 2008).

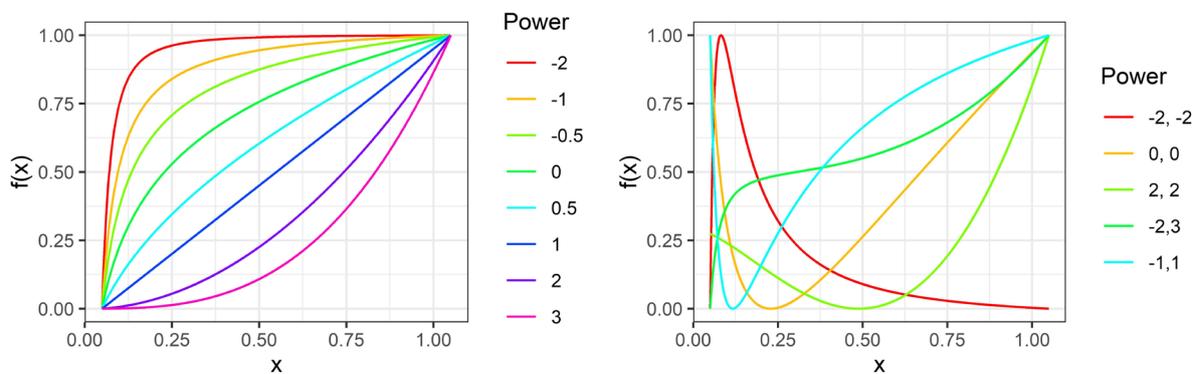

**Figure 1**. Schematic diagram of eight FP1 (left panel) and subset of FP2 (right panel) functions

In the MFP context, the FSP is conducted in a model adjusting for other variables (with their corresponding selected FP functions) currently in the model. The deviance (minus twice the maximized log likelihood) of the null model, the linear model, the best FP1 model and the best FP2 model are compared if FP2 is the most complex function. The procedure starts with a comparison of the best FP2 model with the null model (step 1). If significant, the procedure compares the best FP2 function with the linear model (step 2), and again if significant, the best-fitting FP1 is compared with the best FP2 (step 3). A non-linear FP function is chosen only if it fits the data significantly better than the linear function. If non-linearity is required, a simpler (FP1) function is preferred to a more complex (FP2) function. The use of a closed test procedure ensures that the overall type 1 error rate is close to the nominal significance level (Marcus et al. 1976; Royston and Sauerbrei 2008). For MFP it is important to note that if $\alpha = 1$ for variable selection, then $x$ is "forced" into the model and step 1 is redundant. If the best-fitting FP1 function is linear, step 3 is not required. See section A1 of the Web Appendix for more details on FSP.

## 3 Influential Points and Model Replicability

Influential points are single or pairs (triples) of observations which have an unduly large influence on the selection of an FP function for a particular variable (Royston and Sauerbrei 2007). The leverage of such observations may be high; for example, an FP2 model may be made statistically significant compared with FP1 by a single extreme observation of $x$. This is overfitting and should be avoided because inferences from a model strongly influenced by a single observation are unlikely to be reliable or generalize well to new data. After selecting a function using the FSP, it is important to check whether eliminating any individual observations (or pairs of observations) influences the significance of any of the three FSP tests and thus the selected function.

## 3.1 Identification of Influential Points in Univariable Analysis

### 3.1.1 Diagnostic Plot for Single Points

In accordance with the leave-one-out approach as proposed in the seminal article by Cook (1977) on influential observations, Royston and Sauerbrei (2008) suggested that diagnostic plots be used to identify observations potentially influencing the selection of a function. Successively deleting each single observation from the original dataset, the deviance of the null model, the linear model, and the best-fitting FP1 and FP2 models was stored, and the deviance differences between model pairs were calculated (FP2 vs. null, FP2 vs. linear, and FP2 vs. FP1) and plotted against the deleted observation number or observed variable value. The $\chi_k^2$ critical values with $k$ degrees of freedom and a significance level of α = 0.05, *i.e.*, FP2 vs. null (9.488 for $k = 4$), FP2 vs. linear (7.815 for k = 3), and FP2 vs. FP1 (5.991 for k = 2) were used to decide whether a point was influential or not. For illustration, we also showed corresponding lines for significance level of α=0.01 with critical values of 13.277 ( k = 4), 11.345 (k = 3), and 9.210 (k = 2). Observations which influence the choice of an FP model can be easily observed because their deletion changes the deviance difference, sometimes dramatically compared to the other observations. If the deviance difference is less than the $\chi_k^2$ threshold, there is evidence that the choice of the more complex model depends on this observation or observations, and that a simpler model may be preferred. Since the threshold depends on $\alpha$, an observation may be influential at the 0.05 level but not at the 0.01 level.

### 3.1.2 Diagnostic Plot for Combinations of two or more Points

Royston and Sauerbrei (2008) only discussed the identification of single influential points. The inclusion or exclusion of predictor variables in the model, as well as the functional forms selected, can be influenced by the effect of particular combinations of two or more observations, which can lead to discrepant results. To extend the use of diagnostic plots to detect two or more influential observations, the method described in subsection 3.1.1 was extended by successively deleting a subset of $d$ observations from the original data, which led to $n!/(d!(n-d)!)$ samples. To better understand the effects of influential points, we considered $d = 2$ because higher values can be computationally intensive due to a high number of possible combinations. For each pair $(i,j)$ where $i \neq j$, we constructed samples by removing the $i^{th}$ and $j^{th}$ data points from the original data and fitting fractional polynomial models. Box plots were used to summarize the deviance differences between model pairs for each combination. In total, we had 31,125 replicates generated from a sample size of 250 observations. Pairs containing one specific observation and a subset of the remaining observations are often on opposite sides of the threshold. Boxplots for subgroups of the 31,125 pairs can be used to illustrate the effect of influential pairs. As before, the $\chi_k^2$ threshold was used to determine whether pairs of observations were influential. The approach for searching for triples is straightforward and was not explored here.

### *3.2 Identification of Influential Points in Multivariable Analysis*

Conducting diagnostic analyses for influential points in the multivariable modeling raises additional issues, and we illustrated two approaches. First, we checked for influential points in each covariate using the approach discussed in subsection 3.1.1. Then all observations that were influential for at least one variable were eliminated, and the final MFP model was estimated using the reduced dataset. The second approach started with an MFP analysis of the full data set, followed by a check for influential observations in the selected model. In principle we exchanged the order of checking for influential points and deriving the MFP model. We did not check for IPs in variables excluded from the MFP model.

### *3.2.1 Univariable Analyses to identify IPs followed by MFP on Reduced Data*

Observations identified as IPs for at least one covariate in univariable models were eliminated and an MFP approach was used in data without IPs (the reduced dataset), with the results referred to as IPXu (IP in data X, univariable); later we used IPXm for a multivariable approach to avoid confusion. Although this process of identification uses the univariable analysis of each variable, the observations identified are also likely to influence a joint analysis of the variables. The effects of the observations identified as possible IPs were evaluated by comparing the estimated functions of multivariable models selected on the full data and reduced data.

### *3.2.2 MFP Analysis followed by Check for IPs*

If the underlying model is multivariable, the IPs identified by univariable analysis may differ from those identified by multivariable analysis. Thus, another approach is to perform diagnostic analyses on the MFP model selected using all the data. While adjusting for all other variables in the selected model, the three tests of the FSP for each continuous variable were performed after successively deleting the ith observation (or a pair) as previously conducted in univariable analysis. The FP powers and parameter estimates from the selected MFP model were kept for the adjustment model, whereas Royston and Sauerbrei (2008) kept the power terms but re-estimated regression coefficients in the reduced data. We used the notation IPXm to denote the MFP model in data X after the removal of IPs identified in the multivariable approach.

### *3.3 Model Replicability*

A related issue to influential points is model stability and replicability. In this context, replicability means that the results of fitting MFP models to datasets generated from the same distribution should be identical or nearly identical in terms of variables and functions selected. We demonstrated the replicability of models by selecting MFP models in the three datasets (n = 250) sampled from the ART data: A250 (obs.1-250), B250 (obs. 2001-2250) and C250 (obs.

3001-3250). Influential observations have an impact on the selection of variables and functional forms. Therefore, the functions estimated from the data with and without influential points were compared.

A single model is produced after a model selection procedure is applied to a set of candidate covariates. A very low p-value indicates that a covariate may have a stronger effect and is thus "stable", in the sense that it has a high chance of being selected in similar datasets. For less significant covariates, selection may be more of a matter of chance and the model chosen may be influenced by the characteristics of a small number of observations. If the data is slightly altered, a different model may be selected. Studies assessing the stability of variable selection procedures using bootstrap resampling show that the variables with stronger effects are selected in the vast majority of bootstrap replications, whereas those with weak or "borderline significant" effects may enter the model at random (Sauerbrei and Schumacher 1992, Sauerbrei et al. 2015), and their inclusion can be heavily influenced by influential points.

### *3.4 Influence of Sample Size*

The MFP relies on significance tests for variable and function selection and the detection of non-linear functions requires a large sample size. The smaller the sample size (or in survival analysis, the fewer the number of events), the less likely a test is significant at any given significance level. In FSP, a linear function is the default, and if the sample size is insufficient, a variable may be eliminated or a linear function selected, even if the true function is very different. In the context of variable selection, a range of 10 to 25 observations per variable has been recommended in order to derive suitable models for description (Schumacher et al. 2012; Harrell 2015). Larger sample sizes are usually required for function selection to have sufficient power to reject a linear function in favor of a non-linear function.

When a non-linear function is required, Type II errors (falsely inferring a linear function;

second test of FSP not significant) or even eliminating a variable (first test of FSP not significant) can be a serious problem in smaller samples. The effect of sample size on a model selected was demonstrated using different-sized subsets of the ART data i.e., A125 (obs. 1-125), A250 (obs. 1-250), and A500 (obs. 1 - 500). For relatively large sample sizes (n = 500, about 41 observations per variable), model replicability was investigated by comparing the selected MFP models for datasets A500, B500 (obs. 2001-2500), and C500 (obs. 3001-3500). In all data, IPs were checked and results compared after exclusion of IPs.

## 4. Design of the simulated data

This section introduces the simulated data set used to illustrate the MFP approach and investigate the issues of influential points, model replicability, and sample size. The data are publicly available from the MFP website https://mfp.imbi.uni-freiburg.de/.

In the spirit of plasmode simulations (Gadbury et al. 2008), the ART data set is composed of 5,000 simulated observations that mimic the GBSG breast cancer study in terms of the distribution of predictors and correlation structure (see Appendix A.2.2 in Royston and Sauerbrei 2008). It has a continuous response variable y, and 10 covariates. The covariates include six continuous variables (x1, x3, x5, x6, x7 and x10), two binary variables (x2 and x8) and two 3-level categorical variables (x4 and x9), of which x4 is ordinal and x9 is nominal. For each of x4 and x9 two dummy variables with an ordinal (x4) and a categorical (x9) coding were used. The true model used to generate the ART data was given by

$$y = -4 + 3.5x_1^{0.5} - 0.25x_1 - 0.018x_3 - 0.4x_{4a} + 4x_5^{-0.2} + 0.25\log(x_6 + 1) + 0.4x_8 + 0.021x_{10} + \epsilon$$

where $\epsilon$ is the random noise assumed to be independent and identically distributed $N(0, \sigma^2)$ with $\sigma^2 = 0.49$, resulting in $R^2$ of about 0.50. There are five continuous variables and two categorical variables with an effect on the outcome. The power for variable x5 is not an element of a set of FP functions, and so can only be modelled approximately using the FP approach while a value of 1 was added to variable x6 before logarithm transformation due to 0 values. The contribution of each variable to the model fit was assessed using the percentage reduction in $R^2$. The magnitude of the reduction in $R^2$ is a measure of the importance of a variable (Royston and Sauerbrei 2008). It was clear that variable x5 and x6 were the most important variables, since their removal led to a reduction in $R^2$ of about 56% and 17% respectively, while noise variables had a reduction in $R^2$ of less than 1% (Table A2). The variable x5 relates to the number of positive lymph nodes, a variable known to be the dominating prognostic factor in patients with breast cancer.

Data A250 was used to investigate in details the effects of influential observations in selection of variables and functional forms in univariable and multivariable analysis. Details of the distributions and correlation structure for this subset of the data are presented in section A3 of the Web Appendix. To improve readability, understanding of concepts and results of the investigation for IPs, we used a structured approach to summarize the key issues in a two-part profile for methodological studies (section 2 in Web Appendix).

## 5. Results

### 5.1 Univariable Analysis for Continuous Variables

To illustrate the three steps of FSP, all p-values of the univariable function selection for each continuous predictor in dataset A250 were provided (Table 1). The best FP2 model was compared to the null model, a linear model, and the best FP1 model at $\alpha = 0.05$. Variable x5 had an FP2 (0, 3) function, variable x6 had an FP1 (0) function, and variables x1 and x7 had linear terms, whereas x3 and x10 were not significant.

There are clear discrepancies between the results of selecting a function univariably and the true functions from the multivariable model. Two variables with an effect were not selected (x3, x10), whereas one variable without an effect was selected (x7). The only "correctly" selected power term is FP1(0) for variable x6, but without related parameter estimates, power terms are not informative. One reason for the discrepant findings is the multivariable nature of the true model, which takes into account the effects of other variables in the model while deriving the outcome values. Several variables related to outcome were not included in the univariable models, thus, severe residual confounding occurred (Benedetti and Abrahamowicz 2004; Groenwold et al. 2013). This can be an important reason that univariable relationships sevelely mis-model the true functions. In addition, mis-modelling functions can also be attributed to the effects of IPs, specifically in relatively small sample sizes. It is important to note that if the significance level of 0.01 had been chosen for FSP, an FP1 function would have been selected for x5, the linear functions for x6, and x7 would have been excluded.

**Table 1**: Data A250. Univariable analysis for continuous variables. Column 2 - 4 shows the p-values for different FP tests; Column 5 gives the final FP powers and indicates whether a variable was excluded; the last column shows the FP powers for the true multivariable model used to generate the data.

| Variable | FP2 vs. Null | FP2 vs. Linear | FP2 vs. FP1 | Final selection | True model |
|---|---|---|---|---|---|
| x1 | 0.001 | **0.165** | 0.391 | 1 | 0.5, 1 |
| x3 | **0.856** | 0.765 | 0.659 | out | 1 |
| x5 | 0.000 | 0.000 | 0.033 | 0, 3 | -0.2 |
| x6 | 0.002 | 0.046 | **0.265** | 0 | 0 |
| x7 | 0.012 | **0.449** | 0.678 | 1 | out |
| x10 | **0.135** | 0.314 | 0.295 | out | 1 |

### *5.1.1 Diagnostic Plot for Single Observations*

Diagnostic plots of deviance differences for the three steps of FSP for each observation removed were created to illustrate the three test of FSP and visually examine the data set for the presence of observations that alter the functional form of a selected FP model or selection of variables.

Figure 2 shows the results of the three model comparisons for two variables (x5 and x6) with IPs. For x5, the first two FSP tests were significant, irrespective of the two significance levels. Three IPs (obs. 16, 151, and 175, shown as black dots) were identified as observations which affect the shape of the selected function for variable x5 (top right). If any of these observations were removed, the FP2 vs. FP1 test would be non-significant at the 5% level, resulting in the choice of a simpler FP1 model. Although the values of x5 for observations 16 and 175 were in proximity to other observations, the former had a larger influence on the deviance difference and is thus the first potential candidate to be eliminated from the data. In principle, we could have used a stepwise approach and eliminated one observation at a time (starting with obs. 175 because it had the largest influence or with obs. 151 because it had an outlying value for x5) before repeating the investigation with the remaining 249 observations.

To illustrate a different situation, we present results for variable x6 (lower panel of Figure 2). The first FSP test (FP2 vs. Null) was significant at 0.05 amd 0.01 levels. Several interesting aspects were revealed in the second (FP2 vs. Linear) and third (FP2 vs. FP1) tests. First, both tests were significant at a 0.01 level when obs. 126, was eliminated. This indicated that removing this observation resulted in an FP2 function. Second, the elimination of observations other than 126 cast doubt on the need for a non-linear function because all of the deviance difference values (FP2 vs. Linear) were close to the chi-square critical value at the 0.05 level, with fewer values below the critical value, suggesting that their removal would result in the selection of a linear function.

### *5.1.2 Diagnostic Plot for Combinations of two Observations*

Deletion of pairs of observations to identify possible IPs was also conducted. Figure 3 displays the deviance differences for the last two FSP tests summarized using three groups of boxplots for variables x5 and x6 that had IPs. Group G1 shows the distribution of deviance difference for all 31,125 possible pairs. G2 shows the distribution of pairs with identified IPs, while G3 represents the distribution of pairs with at least one of the IPs excluded.

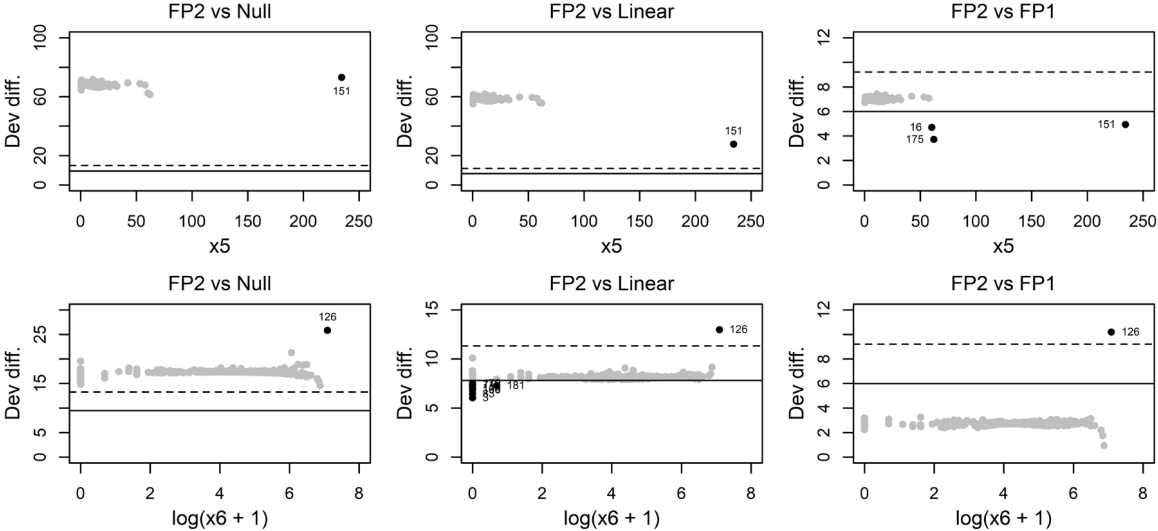

**Figure 2**. Data A250. Plots of deviance differences for each model comparison against observed values for variables x5 and x6. A logarithm scale was used for variable x6 to ease visualization due to extremely large values. Please note that y-axis scales differ. Two threshold values, representing the significance levels $\alpha = 0.05$ and $\alpha = 0.01$, are shown on the plots as horizontal solid and dashed lines, respectively. Please note that the test of FP2 vs linear and FP2 vs FP1 may not be relevant if the test of FP2 vs Null is not significant. Nevertheless, we will always show the full panel.

For variable x5 (top-left panel), two groups of deviance differences were evident, as shown in G1 under the test of FP2 vs. linear. It was clear that obs. 151 was the grouping factor, as any pair with this observation deleted belonged to one group (G3), otherwise, it belonged to the second group (G2), with the two groups being mutually exclusive. The deviance difference was considerably reduced when pairs with obs. 151 were removed, but the test of FP2 vs. linear was still significant, indicating that a non-linear function was needed for x5. Similarly, in the test of FP2 vs. FP1, the groups are separated by the Chi-square threshold at 5.991, indicating that the elimination of at least one of the observations 16, 151 or 175 (group G3) led to the non-significance of the test in most cases, resulting in the selection of an FP1 function. The inclusion of at least one of these three observations (group G2) led to an FP2 function for the significance level of 0.05. Deletion of pair (126, 151) resulted in the selection of an FP2 (-0.5, 3) function instead of a simpler FP1. Further scrutiny on the functional plot (bottom-left panel of Figure 3) after the deletion of pairs (126, 151) revealed that obs. 16 and 175 were the main causes of an FP2 function. This confirms that the three observations (16, 151, and 175) were indeed influential. Deletion of these three observations produced a simpler FP1 (-0.5) function, pointing out that the complex FP2 function was not required.

For variable x6, the test of FP2 vs. FP1 (Figure 2, bottom–right) identified two groups. The second group (G2) contained all pairs with influential obs. 126. Its presence in the data resulted in the selection of an FP1 function, while its deletion resulted in an FP2 function at 0.05 level (G3). The deletion of a pair (14, 126) revealed that another observation number 14, which was not influential in single-case deletion, was influential. This explains why an FP2 function was selected when obs. 126 was deleted (Figure 3, top–right). After removal of the two influential points (14 and 126), an FP1 (-0.5) function was selected (Figure 3, bottom–right panel). Hence, it was sufficient to describe x6 using a simpler FP1 function rather than an FP2 function.

### *5.1.3 Plot of Functions*

Figure 3 displays the functional forms of variables x5 (top-left panel) and x6 (top-right panel) before and after IP removal. There were no IPs found for the other continuous variables (x1, x3, x7, and x10). For x5, the true function FP1 (-0.2) was quite similar to the FP2 (0, 3) function from all the data up to about x5 = 50. Thereafter, there was a huge deviation due to the influence of obs. 151. The FP1 (-0.5) function obtained by omitting observations 16, 151, and 175 was a better approximation of the true function than the FP2 function estimated from all the data. The larger uncertainty (wider 95% point-wise confidence interval) towards the right end is a result of fewer observations with values of x5 larger than 50. It is important to note that the uncertainty of the function is underestimated because the function was derived data-dependently, an aspect ignored here. Furthermore, the estimated function refers to a univariable model, whereas the data were generated using a multivariable model with some correlated covariates.

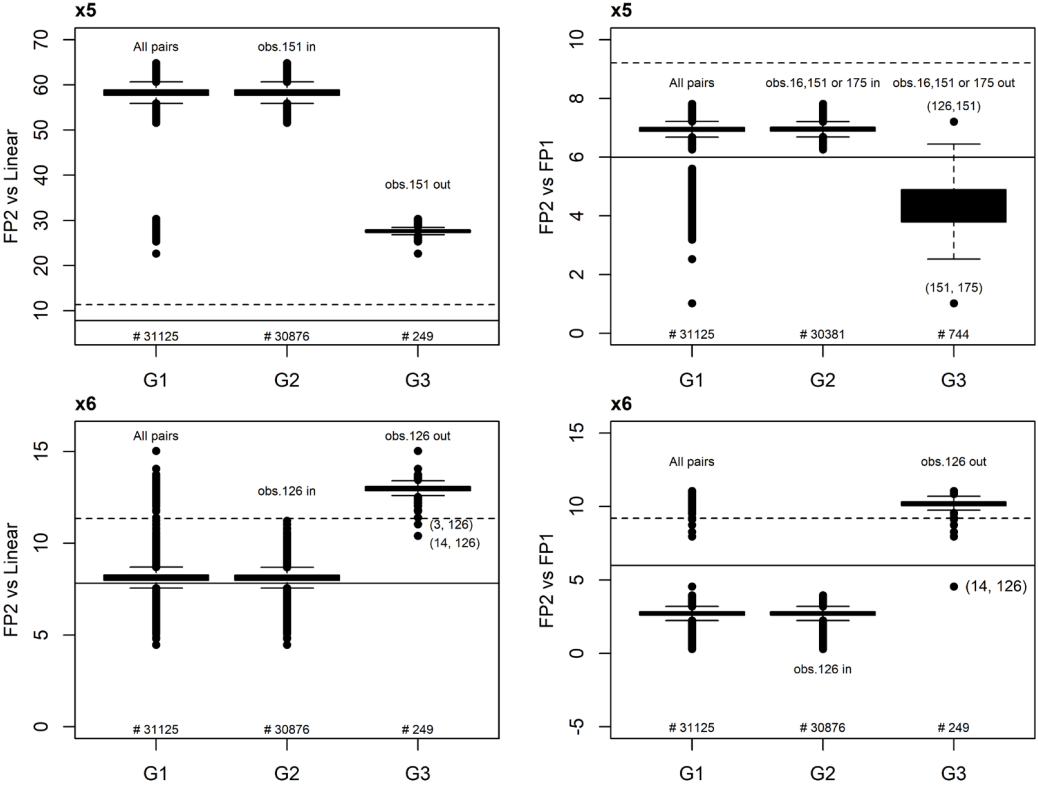

**Figure 3** Data A250. Detection of influential points in variable $x_5$ and $x_6$ by deleting pairs of observations. The dashed and solid horizontal lines denotes the FSP test at 0.01 and 0.05 level respectively. Influential observations are highlighted on the graph.

The true and selected function for variable x6 with all the data was slightly different even though both functions were FP1(0) (top-right panel). The difference was caused by true and estimated coefficients ($\beta_{true} = 0.25$ and $\hat{\beta}_{estimated} = 0.15$) as well as the effects of influential obs. 126. Deletion of obs. 126 resulted in an FP2 (-1, 3) function, but closer inspection revealed that the data might contain other influential observations (e.g., obs. 14 or 218).

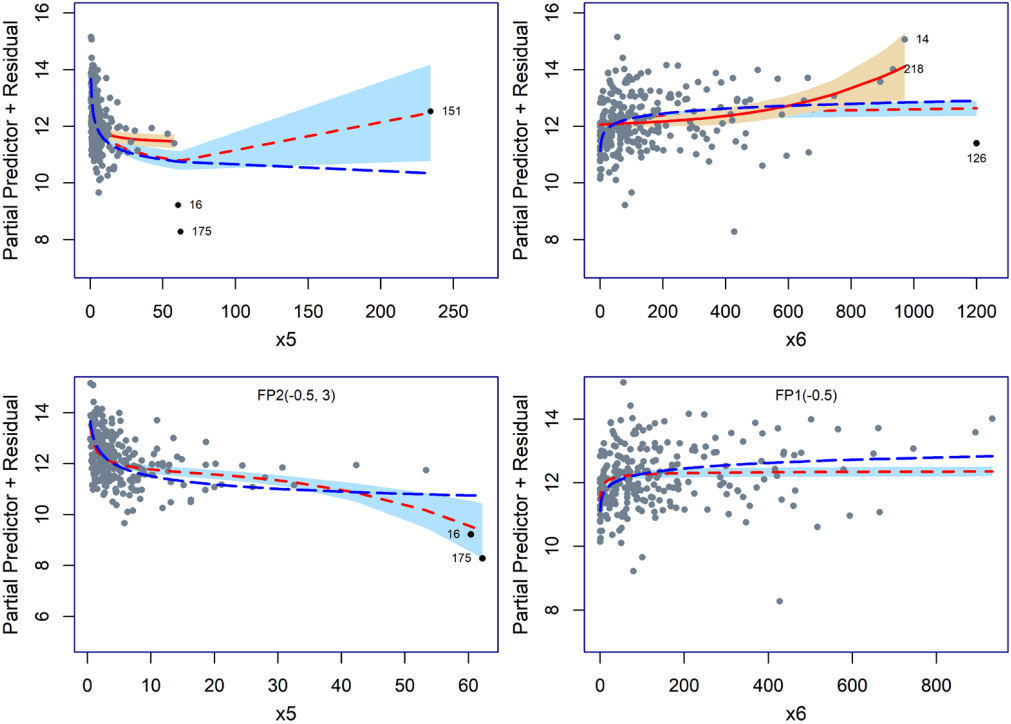

**Figure 4.** Data A250. Functional forms of variables x5 and x6. Top: the estimate of the functional form from complete data (red, short-dashed), data without IPs identified using the L-1 approach (solid line) and true function (blue, long-dashed line). Bottom: the estimate of the functional forms of variables x5 (left) and x6 (right) after the removal of observation (126, 151) and (14, 126), respectively. Please note the different scales.

### *5.1.4 Investigation of Function Replicability*

The replicability of the selected univariable functions was investigated across three data sets (A250, B250, and C250). The functional forms of continuous variables were compared before and after IPs were removed as shown in Figure 5 which is based on the results of Table 2. The graph of variable x5 (top-middle panel) demonstrates how an IP can lead to unnecessary complex function. When the IP was removed (bottom-middle panel), the functional form of variable x5 was quite similar to the true function. A linear function of variable x1 did not correspond to the true FP2 function (bottom left). For variable x6, the functional forms were similar to the true function after IPs were removed as expected since this variable had a strong effect and the correlation in the data was low. These findings indicate that function replicability is influenced by both sample size and influential points. More information on identifying influential points in data B250 and C250 can be found in section 5 of the Web Appendix.

**Table 2**: Data A250, B250 and C250. Univariable analysis for continuous variables."All data" and "all data-IPs" refers to FP powers obtained with complete data and after removing IPs, respectively. Variable (a, b, c) refers to the total number of IPs for each variable in each dataset; where a, b and c stands for A250, B250 and C250 respectively. '=' denotes same power term selected

| Variable(a,b, c) | A250 | | B250 | | C250 | | True |
|---|---|---|---|---|---|---|---|
| | All data | All data-IPs | All data | All data-IPs | All data | All data-IPs | |
| x1(0, 0, 1) | 1 | = | 1 | = | 3, 3 | 1 | 0.5, 1 |
| x3(0, 0, 0) | Out | = | Out | = | Out | = | 1 |
| x5(3, 0, 2) | 0, 3 | -0.5 | 0 | = | 0 | 0.5, 0.5 | -0.2 |
| x6(2, 0, 1) | 0 | -0.5 | -0.5 | = | 1 | 0.5 | 0 |
| x7(0, 3, 0) | 1 | = | Out | 0 | 0 | = | Out |
| x10(0, 0, 0) | Out | = | Out | = | 1 | = | 1 |

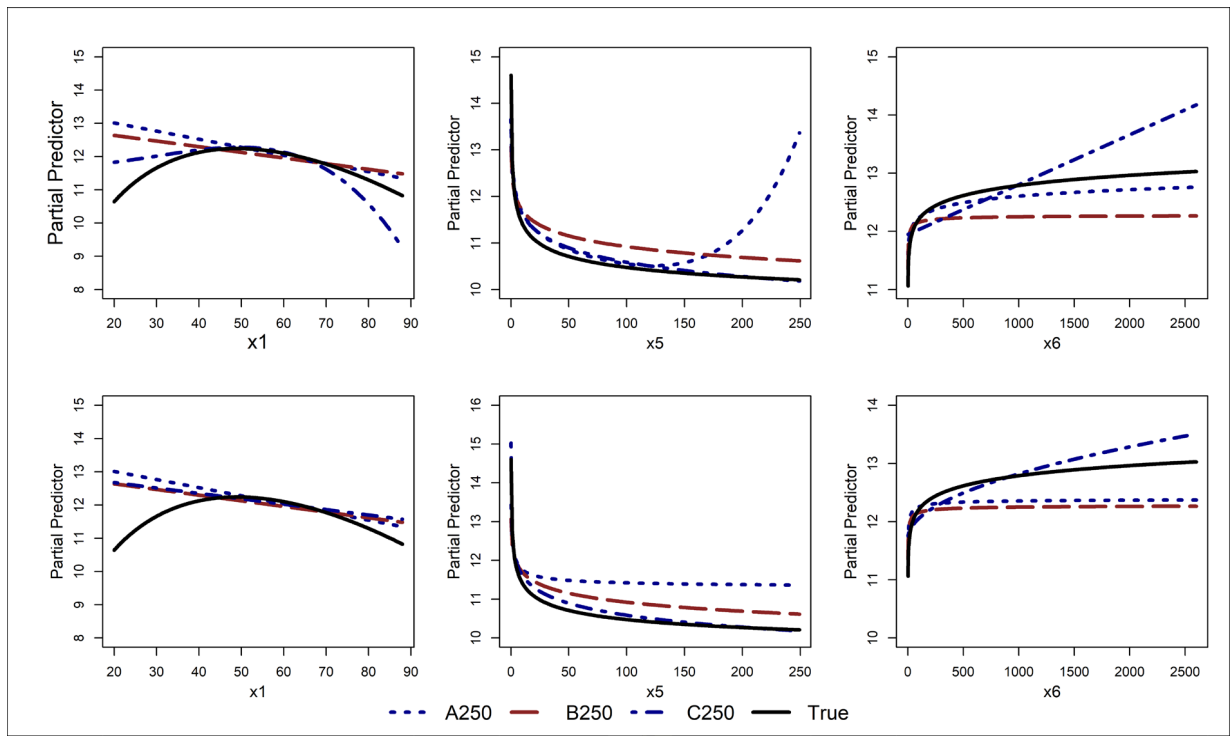

**Figure 5** Data A250, B250 and C250. Functional forms of continuous variables in univariable analysis for x1, x5, and x6 that were selected in three datasets. Variable x10 was only selected in C250 and had a linear function, hence its plot is not provided. The upper panel shows the plots from complete data, while the lower panel shows the plots after the removal of IPs.

## 5.2 Multivariable Analysis – Effect of Influential Points

### 5.2.1 Elimination of Influential Points identified in Univariable Analysis

MFP analyses were run to generate multivariable models for data A250, B250 and C250 before the IPs were deleted. The selected models are displayed in Table 3, in the column labelled "all". Next, all the IPs identified in univariable analyses for the six continuous variables were deleted and an MFP model was fitted, the results of which are displayed in the column labelled "IPXu". Finally, the column labelled "IPXm" presents the MFP model selected after deleting IPs that were identified in the diagnostic analysis of the multivariable model.

In the univariable analysis, a total of 5, 3, and 4 influential observations were identified in A250, B250, and C250, respectively. Deleting these observations resulted in the selection of variables similar to the model fitted to the full data. However, in data A250, a simpler FP1 (-0.5) function was estimated for variable x5 after deleting IPs rather than an FP2 (0, 3) function from complete data. In data set B250, different powers of FP2 functions were also estimated for variable x1. Compared to the results from the univariate investigations (Table 2), several functions differ substantially. For x1 a linear function was selected in B250 whereas an FP2 is selected with the multivariable approach (all and IPBu). In A250 x3 was not significant in the univariate analyses but was included with a linear function in the multivariable case.

### *5.2.2 Diagnostic Analyses on Multivariable Model*

Diagnostic analyses were performed on the selected multivariable model (column "all" in Table 3) as a second way to check for IPs in a multivariable context. The IP investigation for dataset A250 is described in this section, while the IP investigations for datasets B250 and C250 were described in section A4 of the Web Appendix.

### *Identification of Influential Points in Data set A250*

In leave-one-out approach (Figure A4, Web Appendix), obs. 175 was found to influence the functional form of variable x5 at the 0.05 level. Its removal turned an FP2 (0, 3) function into an FP1 (-0.5) function. In the leave-two-out approach, IPs were found in variables x5 and x10. For variable x5, deletion of any pair with obs. 175 rendered the test of FP2 vs. FP1 non-significant except when two pairs (37, 175) and (151, 175) were deleted (Figure A5). An inspection of the functional forms (Figure A5, Web Appendix) revealed that when a pair (37, 175) was deleted, an FP2 function was estimated because of the effects of obs. 151 that was still in the data. Similarly, when a pair (151, 175) was deleted, an FP2 function was driven by obs. 37. As such, observations 37, 151, and 175 were indeed influential in variable x5. An easy and informal way to check for the three IPs simultaneously is by deleting three observations at a time instead of pairs. Only two observations, 151 and 175, were influential in both univariable and multivariable checks for IP. For variable x10, deleting two pairs (37, 76) and (74, 76) rendered the test of FP2 vs. linear significant, implying that observations 37, 74, and 76 were IPs. Deleting any of the pairs led to an FP1 function. In total, five IPs (37, 74, 76, 151, and 175) were identified in A250 as presented in Table 3.

**Table 3:** Data A250, B250 and C250. Selected MFP models with complete data ("all") and after removal of IPs identified from the univariable (IPXu) and multivariable (IPXm) diagnostic analyses. The number of IPs identified univariable and multivariable, respectively, are shown in parentheses .'=' is used if the power selected agreed to the power from all data.

| Type | Variable | Dataset A250 (5, 5) | | | Dataset B250 (3, 3) | | | Dataset C250 (4, 6) | | | True model |
|------|----------|------|------|------|------|------|------|------|------|------|------|
|      |          | all  | IPAu | IPAm | all  | IPBu | IPBm | all  | IPCu | IPCm |      |
| Cont. | x1 | 1 | = | = | -1,3 | 0,3 | 1 | Out | = | = | 0.5, 1 |
|       | x3 | 1 | = | = | out | = | = | Out | = | = | 1 |
|       | x5 | 0, 3 | -0.5 | 0 | 0 | = | = | -0.5 | = | = | -0.2 |
|       | x6 | 0 | = | = | -0.5 | = | = | 0 | = | = | 0 |
|       | x7 | out | = | = | out | = | = | Out | = | = | out |
|       | x10 | 1 | = | 3 | out | = | = | 1 | = | out | 1 |
| Cat.. | x9a | out | = | = | in | = | = | Out | = | = | out |
|       | x9b | out | = | = | out | = | = | Out | = | = | out |
| Bi    | x2 | out | = | = | out | = | = | In | = | = | out |
|       | x4a | in | = | = | in | = | = | In | = | = | in |
|       | x4b | out | = | = | out | = | = | Out | = | = | out |
|       | x8 | in | = | = | out | = | in | In | = | = | in |

Cont, Cat and Bi denotes continuous, categorical and binary variables respectively

Table 3 compares models from complete data ("all") and after eliminating IPs (IPAu and IPAm) in the three datasets with a sample size of 250. . Elimination of IPs had an influence on some of the selected functions (x1 in B250, x5 in A250, and x10 in A250 and C250). IPs had also an influence on the selection of the binary variable x8 in B250. In particular, in A250 an FP2 (0, 3) function was estimated for variable x5 due to the effects of IPs and a satisfactory function is FP1 (0) which was quite similar to the true function (Figure 6). However, elimination of IPs may also result in the selection of a non-linear function instead of a linear function (x10 in A250).

More important is the comparison of the selected models with the true model. Concerning the inclusion of binary and categorical variables, we observed full agreement in A250, a difference for x2 in C250, and some differences in B250. Concerning power terms of functions, we observed good agreement for x6 and x7 (was always out), and non-linear functions selected for x5 in all analyses. Several disagreements were observed in other variables. In specific, for x1, where FP2 was the true function, but the variable was excluded in C250 and a linear function was estimated in A250, a strong indication that the power was insufficient to identify the non-linear effect. A larger sample size seems to be needed.

Figure 6 compares the functional forms of continuous variables from three datasets with and without IPs. The true function for variable x1 was an FP2, which was well approximated by data B250 before the removal of IPs but elimination of IPs resulted in the selection of a linear function which is far away from the true effect of x1. The true and estimated functions for variables x5 and x6 were nearly identical when IPs were removed.

### *5.3 Sample Size and its Effect on Identifiability of the True Model*

To evaluate the effect of the sample size on the identifiability of the models, we compared models derived with different sample sizes and also after IPs were deleted. Univariable and multivariable approaches were used to check for IPs.

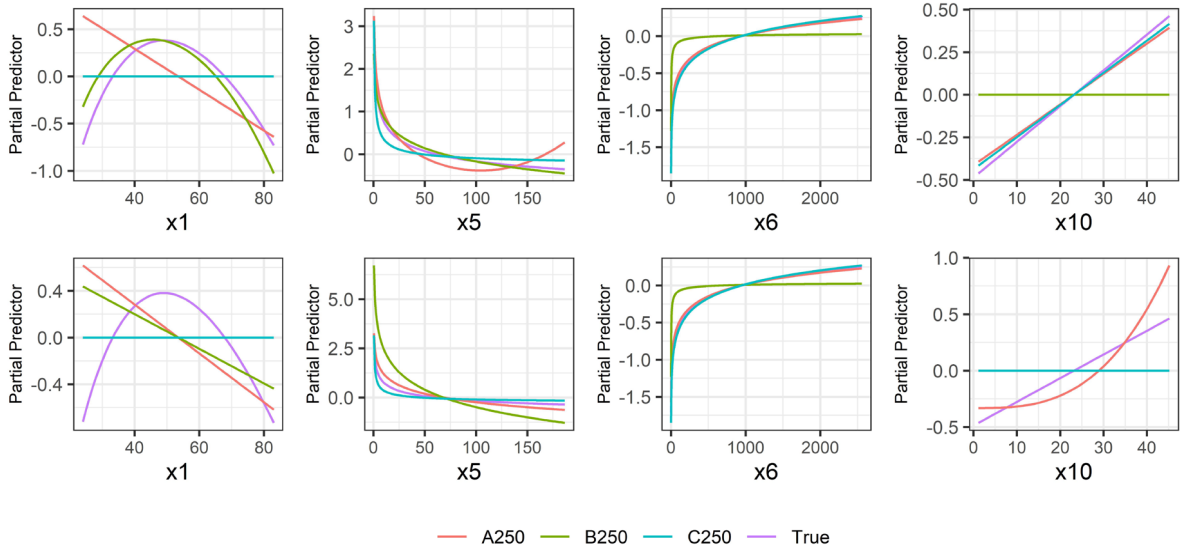

**Figure 6** Data A250, B250 and C250. Functional forms of continuous variables for the selected MFP models (see Table 3). The upper panel shows the plots from complete data, while the lower panel shows the plots after the removal of IPs. The horizontal line indicates that no variable was selected. Not shown are x3 (linear in true and A250, out in B250 and C250) and x7 (true out and never selected).

*Small to relatively large dataset*

Table 4 summarizes the power terms of the nine models selected from small to relatively large datasets, while Figure 7 shows related functions for data without IPs (i.e., IPAm).

Multivariable analysis of the complete data set A125 led to the selection of only three variables: x5, x6 and x8 due to low power for selecting variables with moderate effects (Table 4). Even though the sample size was relatively small, non-linearity of x5 and x6, the two variables with a stronger effect (see Table A2), was identified. The removal of three IPs (obs.14, 16 and 105) that were identified in the univariable approach led to the inclusion of variable x3 and changed the FP1 power term for variable x5. No IPs were found in the diagnostic analysis of the multivariable model of data set A125. Compared to the true model, the main differences in selected MFP models were the elimination of x1, x10, and x4a, while x3 was only included with the IPu approach. These results illustrate that the sample size of 125 was much too low to select a suitable MFP model.

The results for the sample size of n = 250 were much closer to the true model since variables x1 (although only linear), x3, x10, and x4a were included in the model. The elimination of five influential points did not affect the selection of variables but changed some of the power terms of continuous variables. For n = 500, the selected MFP models agreed well to the true model. Selected functions for x1 (Figure 7) best illustrate the significant impact of sample size. The variable was eliminated when n = 125, a linear function was selected when n = 250, and an FP2 function that was close to the true function was selected when n = 500.

**Table 4**: Data A125, A250 and A500. Selected functions from MFP models with all data ("all") and after removal of IPs identified from the univariable ("IPAu") and multivariable ("IPAm") diagnostic analyses. The number of IPs identified in each respective analysis is shown in parentheses next to the name of the data.

| Type | Variable | Data A125 (3, 0) | | | Data A250 (5, 5) | | | Data A500 (6, 1) | | | True model |
|---|---|---|---|---|---|---|---|---|---|---|---|
| | | all | IPAu | IPAm | all | IPAu | IPAm | all | IPAu | IPAm | |
| Cont. | x1 | out | = | = | 1 | = | = | 0.5, | = | = | 0.5, 1 |
| | x3 | out | 1 | out | 1 | = | = | 1 | = | = | 1 |
| | x5 | 0 | -0.5 | 0 | 0, 3 | -0.5 | 0 | 0, 3 | -1, 0 | 0 | -0.2 |
| | x6 | 0 | = | = | 0 | = | = | 0 | = | = | 0 |
| | x7 | out | = | = | out | = | = | out | = | = | Out |
| | x10 | out | = | = | 1 | = | 3 | 1 | = | = | 1 |
| Cat. | x9a | out | = | = | out | = | = | out | = | = | Out |
| | X9b | out | | | out | = | = | in | = | out | Out |
| Bi. | x2 | out | = | = | out | = | = | out | = | = | Out |
| | x4a | out | = | = | in | = | = | in | = | = | In |
| | x4b | out | = | = | out | = | = | out | = | in | Out |
| | x8 | in | = | = | in | = | = | in | = | = | In |

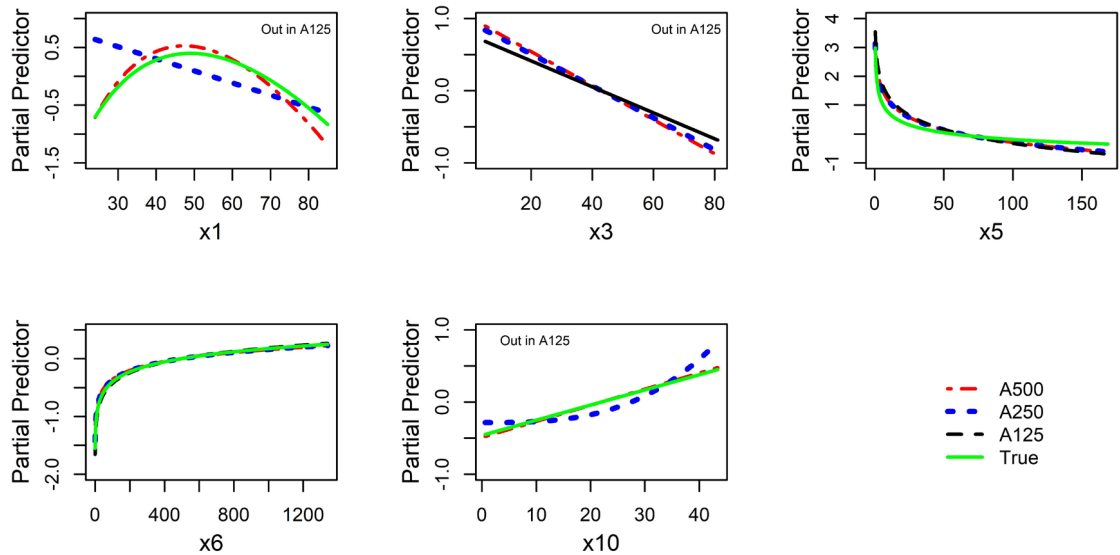

**Figure 7** Data A125, A250 and A500. Functional forms of continuous variables after elimination of influential points identified in multivariable model (results of IPAm in Table 4). Variables x1, x3 and x10 were not selected in A125.

*Relatively large dataset*

Results for three relatively large datasets (A500, B500, and C500) were summarized in subsection 4.3 of the Web Appendix. Influential points had some effects on the power terms chosen, and binary variables were not always correctly included. Figure 8 shows the estimated functions (after deletion of IPs) for the five continuous variables that had an effect on the outcome. In C500, a non-linear function was estimated for variable x10 instead of the correct linear function but otherwise agreement is excellent. Identification and elimination of IPs improved the selected function for x5 (FP2 in all data, FP1 after removal of IPs), changed the selection of x9b and x4b in A500 but otherwise the effect was negligible in the three data sets.

Generally, with large sample sizes and removal of influential points, variables selected and the estimated functional forms were good approximations of the true model.

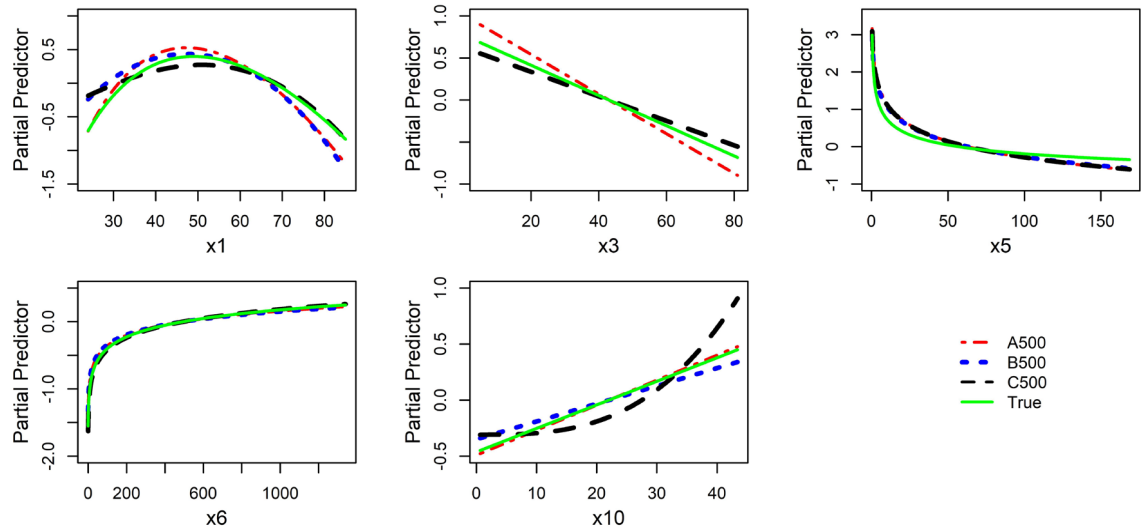

**Figure 8** Data A500, B500 and C500. The plots were created after removing the influential points identified in the multivariable model (results of IPAm in Table A5). Variable x7, which was irrelevant in the true model, was not selected in each data set, so it was not plotted.

## 6. Discussion

In areas of science in which empirical data are analyzed, various types of regression models are derived for prediction, description, and explanation (Shmueli 2010). In medicine, continuous measurements such as age and weight are often used to assess risk, predict an outcome, or select a therapy. Background knowledge or the type of question should strongly influence how continuous variables are used. However, knowledge is often insufficient and the analyst needs to decide how to handle continuous variables, a very difficult issue in the context of multivariable analysis when the selection of the functional form of a continuous variable needs to be combined with the selection of variables which have an influence on the outcome.

Concerning continuous variables, categorization or the assumption of a linear effect are still the most popular approaches (Shaw et al. 2018), despite many well-known weaknesses (Greenland 1995; Altman 1994; Royston 2006; Sauerbrei et al. 2020). This unfortunate situation is partly caused by lack of guidance for the selection of variables and modelling of continuous variables. Sauerbrei et al. (2020) described and discussed the fractional polynomial and spline-based approaches in an overview paper of topic group 2 "Selection of variables and functional forms in multivariable analysis" of the STRengthening Analytical Thinking for Observational Studies (STRATOS) initiative (Sauerbrei et al. 2014). Various spline-based approaches have been proposed and an overview of the most widely used spline-based techniques and their implementation in R software is given in Perperoglou et al. (2019). The authors illustrated some challenges that an analyst face when working with continuous variables using a series of simple scenarios of univariable data. They concluded that an

> *experienced user will know how to obtain a reasonable outcome, regardless of the type of spline used. However, many analysts do not have sufficient knowledge to use these powerful tools adequately and will need more guidance.*

Univariable analysis was the emphasis of this overview. A brief overview of spline-based techniques for multivariable model building was given in Sauerbrei et al. (2020). While FPs are global functions, splines are much more flexible and can also estimate local effects. However, that comes at the price of more function instability and uncertainty (Sauerbrei et al. 2007). Furthermore, local features may be identified by a systematic check of residuals, and statistically significant local polynomials can be parsimoniously added (Binder and Sauerbrei 2010). Results of MFP and spline-based approaches were compared in several examples (Sauerbrei et al. 2007, Royston and Sauerbrei 2008), and a simulation study (Binder et al. 2013), but it is obvious that more comparisons of spline procedures in both univariable and multivariable contexts and comparisons to MFP are needed.

In contrast to the spline approaches, the MFP procedure is a well-defined pragmatic approach. Deriving suitable models for description is the main aim, and the two significance levels for the BE and FSP parts are the key tuning parameters. Using simulated data, we illustrated all steps of the procedure and the importance of checking whether influential points affect (strongly) the selected model with the potential consequence of (severe) errors in variables or functional forms selected. IPs can also have a strong effect on model (in-)stability (Royston and Sauerbrei 2004). Leave-one-out and leave-M-out are simple and helpful techniques for the identification of influential points which can be easily understood by most analysts with at least some background in regression modelling. It is important to check each multivariable model that includes continuous variables for potential IPs. Here, we eliminated identified IPs, but other options may be preferable in real data.

The effects of sample size on MFP models were illustrated in datasets A125, A250, and A500 (Table 4 and Figure 7). We observed that MFP models derived from a relatively small sample size (A125) deviated severely from the underlying true model since some relevant variables were excluded and linear functions were estimated for some continuous variables instead of nonlinear functions, probably due to low power to detect nonlinearity (Royston and Sauerbrei 2008). We also observed that an MFP can detect stronger nonlinear functions in small sample sizes (e.g., variables x5 and x6). As the sample size increased (A500), the performance of MFP improved drastically since important variables were correctly selected and nonlinear functions (e.g., x1 and x3) were identified. In addition, all models derived with a relatively large sample size (500 observations, 12 variables, about 42 observations per variable) and IPs eliminated were similar to the true model as shown in Table A5 and Figure 8. These results indicate that with about 50 or more observations per variable, it may be possible to derive suitable descriptive models for studies with several variables ranging from about 5 to 30. In our simulated data, we had six continuous and six binary variables.

The results of the function selection procedure can be driven by influential points. For instance, the estimated functional form for variable x5 (Figure 4) from the complete data with IP is a non-monotonic FP2 function instead of a monotonic FP1 function. Similar results were observed in the case study where an FP2 function was estimated for the variable abdomen instead of a linear function (Table A6). These results indicate that the data analyst needs to use the algorithm carefully while selecting the functional forms of continuous variables since, in some instances, a simple function may suffice instead of a complex function driven by IPs. Plots of deviance differences for variables x5 and x6 (Figures 2 and 3) illustrate that such additional investigation can support the final decision for a model, e.g., we might prefer a simpler model despite a (just) significant result for the more complex model. Comparisons of two competing functions (e.g., linear versus best FP1) may show that the difference is small and subject matter knowledge or practical usefulness may be used as a criteria for the final selection.

As often done, we started with the investigation of one variable, while our outcome was created according to a multivariable process. Such marginal investigation may be misleading, and researchers may prefer to derive a multivariable model and check whether single points have a severe influence on the model selected. In several datasets, we conducted such an approach and found some differences in potential IPs identified. We did not check whether variables eliminated by MFP would have been included if we had eliminated single observations from the data set. In real data, we would recommend that. If a single continuous variable is of main interest (e.g., a continuous risk factor in epidemiology), it is straightforward to use our "univariable" investigations, adjusted for relevant confounders, to check whether single points drive the selected function for this variable.

# 7. Conclusions

Variable selection by using backward elimination and the function selection procedure are easily understood and can be suitably used by non-experts. It is obvious that the sample size needs to be sufficient and aspects of model criticism should be standard for each derived multivariable model. We concentrated on the importance of influential points, but further aspects (e.g., residual plots) are also relevant. Some issues are discussed in chapters 5, 6, and 10 in Royston and Sauerbrei (2008), and more issues can be found on the MFP website. If the effect of continuous variables needs to be investigated in the context of a multivariable regression model, recommendations for practice were proposed under several assumptions (Sauerbrei et al. 2007; Royston and Sauerbrei, 2008, Chapter 12.2).

**Acknowledgement**

Part of the paper is based on chapter 10 of the book by Royston and Sauerbrei (2008), and we used the data simulated for this chapter. Here we repeated and extended some of the analyses and discussed a real example in detail. Patrick Royston would have been an obvious co-author based on the joint work on the book, but he felt that his contribution to the paper was insignificant and that he did not meet the standards for authorship. Thank you so much, Patrick, for all your efforts on MFP methodology and for close friendship with WS. We appreciate Yessica Fermin's help with a much earlier version of the paper as well as Jakob Moeller and Sarah Hag-Yahia for administrative assistance. This work was supported by the German Research Foundation (DFG) to WS under grant SA580/10-1.

The authors report that there are no competing interests to declare.

# Appendix

# Effects of Influential Points and Sample Size on the Selection and Replicability of Multivariable Fractional Polynomial Models

Willi Sauerbrei[1*], Edwin Kipruto[1*], James Balmford[1+],

[1]Institute of Medical Biometry and Statistics, Faculty of Medicine and Medical Center - University of Freiburg, Germany

*Joint first authorship; + deceased

## 1 Function Selection Procedure

As a first step in the function selection procedure (FSP), 44 regression models are fitted, each with a different FP1 or FP2 function applied to the continuous covariate in question. The deviance of each fitted model is determined, and using deviance differences the best-fitting FP1 and FP2 functions are selected.

The FSP then continues in 1, 2 or 3 steps as follows:

1. Test the best FP2 model for $x$ at the α significance level against the null model using a test with four degrees of freedom. If the test of deviance difference is not significant, it should be stopped, and concluded that the effect of x is not significant at the α level, implying that the variable should not be included in the model. Otherwise it should be continued.

2. Test the best FP2 for $x$ against the linear function at the α level using a test with three degrees of freedom. If the test is not significant, stop, and conclude that $x$ is best modelled by a linear relationship with the response. Otherwise continue.

3. Test the best FP2 for $x$ against the best FP1 at the α level using a test with two degrees of freedom. If the test is not significant, the final model is the best FP1, otherwise the final model is the best FP2. End of procedure.

## 2 Two-Part Structured Approach

To improve readability, understanding of concepts and results of the investigation for IPs, we used a structured approach to summarize the key issues in a two-part profile for methodological studies. Part A follows the ADEMP structure that was recently proposed by Morris et al. (2019) to improve understanding and interpretation of simulation studies. Part B provides a summary of all analyses as proposed in the explanation and elaboration paper of the REMARK reporting guidelines for tumor markers and extended in a case study on the development and assessment of prediction models (Altman et al. 2012; Winzer et al. 2016). Table A1 provides an overview of the main aim of the study, data used, and analyses conducted. We analyzed seven subsets of an artificial data set and illustrated the methods in a real data set. As proposed in the REMARK profile, the second part provides an overview of all analyses conducted and distinguishes between data description (D), analyses (A), and presentations (P). Here, investigation of model assumptions was not carried out, an important issue that should be conducted in general. In total, 21 analyses were conducted, often with parts a and b. This notation was used to show that the same analyses were conducted with leave-one-out or leave-two-out, in different datasets or in univariable or multivariable analyses. The results/remarks section points to the specific table or figure.

**Table A1:** MethProf-simu profile giving an overview of the aims, data, estimand or target of analysis, methods and performance measures (ADEMP structure) in part A. All analyses are listed in part B, categorized into analysis (A), presentation (P) and description of data (D).

**Part A ADEMP structure**

| **A**ims | - To investigate whether IPs exist and have an influence on the selected FP functions in univariable and multivariable fractional polynomial (MFP) models. |
| --- | --- |
| | - To investigate replicability of MFP models |
| | - To investigate the effects of sample size on selecting a 'suitable' MFP model |
| **D**ata | - We used the ART data set published on the website http://mfp.imbi.uni-freiburg.de/book. The data simulated 5000 patients in a normal-error regression setting with 10 predictors; six of which are continuous ($x_1$, $x_3$, $x_5$-$x_7$, $x_{10}$), two are binary ($x_2$, $x_8$) and one each ($x_4$ and $x_9$) are ordinal and nominal respectively. The design was influenced by a real Germany breast cancer study. |
| | - We considered seven subsets of the data with different sample sizes abbreviated as: **A125**(obs.1-125), **A250**(obs.1-250), **A500**(obs.1-500), **B250**(obs.2001-2250), **B500**(obs.2001-2500), **C250**(obs.3001-3250) and **C500**(obs.3001-3500) |

|  |  | - We also analysed the body fat data (Johnson 1996) where the outcome and predictors ($p = 13$) were all continuous and the sample size $n = 252$ observations |
| --- | --- | --- |
| **E**stimand/target of analysis | | The targets: (i) investigate effects of single and pairs of observations on the deviance of models relevant for the FSP strategy (best FP2 vs. Null, best FP2 vs. Linear, best FP2 vs. best FP1), (ii) assess replicability by comparing models selected in different data sets with the same sample size and derived from the same true model and (iii) to compare models derived with different sample sizes. |
| **M**ethods | | - Function selection procedure (FSP) to select FP functions<br>- Leave-d-out-approach (d =1 and 2) to identify IPs<br>- In multivariable analyses we used two approaches<br>i) Eliminate 'univariable' IPs followed by MFP on reduced data<br>ii) MFP on all data followed by check for IPs in selected model<br>- The MFP ($\alpha_1, \alpha_2$) procedure was used with $\alpha_1 = \alpha_2 = 0.05$ for function and variable selection. Critical value for $\alpha_2 = 0.01$ was also shown to discuss some issues. |
| **P**erformance measures | | - Comparisons of the true functions and the true models with functions derived with FSP and models derived with MFP for all data and after elimination of IPs.<br><br>- Use of graphical comparisons for functions. Comparisons are conducted for different data with the same sample sizes and for varying sample sizes. |

## Part B Overview of analyses and presentations

| Analysis | Dataset | Variables considered | Remarks/results |
| --- | --- | --- | --- |
| **P1**: FP functions | NA | NA | **Fig 1.** Schematic diagram of eight FP1 and five FP2 functions. |
| **D1**: description of data | ART | y, x1-x10 | **Table A2** Variable importance based on $R^2$ |
| **D2**: description of data | A250 | y, x1-x10 as x6 has values '0' we use x6+1 | **Table A3**. Distribution of variables; x5, x6 and x7 have high kurtosis **Table A4**. Correlation structure. Larger difference between Pearson and Spearman presented |
| **A1**: Function Selection Procedure (FSP); Univariable analysis | A250 | y, x1, x3, x5-x7, x10 | **Table 1**. FSP (0.05) x5 and x6 were non-linear |
| **A2a**: check for IPs in univariable analysis; (L-1) | A250 | y, x1, x3, x5-x7, x10 | **Fig 2.** x5 and x6 had IPs. Data not shown (**DNS**) for Investigation of IPs in x1, x3, x7, and x10. No IPs found. |
| **A2b**: check for IPs in univariable analysis; (L-2) | A250 | y, x1, x3, x5-x7, x10 | **Fig 3**: x5 and x6 had IPs. **DNS**: no IPs identified in other variables. |
| **P2**: effects of IPs identified in A2 on functional forms | A250 | x5, x6 | **Fig 4**. Functional forms for x5 and x6 with and without IPs identified in **A2a** and **A2b** |
| **A3a, A3b** as in A2a, A2b | B250 | y, x1, x3, x5-x7, x10 | **DNS**. IPs were only found in x7 |
| **A4a, A4b** as in A2a, A2b | C250 | y, x1, x3, x5-x7, x10 | **DNS**. IPs were found in x1, x5, x6 and x7 |
| **P3**: effects of IPs identified in A4 on model fit | C250 | x5, x6 | **Fig A1** and **A2**. Two and one IPs identified in x5 and x6 respectively, confirmed by residual plots. |
| **P4**: effects of IPs identified in A4a and A4b on functional forms | C250 | x7 | **Fig A3**. IPs introduced a hook in the function |
| **A5**: functions selected in univariable analysis; full data and data without IPs identified in **A2**, **A3** and **A4** | 250 (A/B/C) | y, x1, x3, x5-x7, x10 | **Table 2**. x5 and x6 were non-linear in all data sets after IPs were removed. x1 non-linear in C250 |
| **P5:** presentation of functions selected in A5 | 250 (A/B/C) | y, x1, x5, x6 | **Fig 5.** Similar functions estimated for x1, x5 and x6 when IPs removed. Low power for detecting non-linearity in x1 |
| **A6:** MFP models in full data and in data without IPs | 250 (A/B/C) | y, x1-x10 | **Table 3.** Investigate replicability of MFP models in too small datasets. Several deviations from the true model (See Fig 6). For effect of sample size see Tab 4, Tab A5 and Fig 8. |
| **A7a, A7b:** check for IPs in selected MFP model (see A6); L-1 and L-2 | A250 | **y, x1, x3, x5, x6, x10** | **Fig A4.** IPs only identified in x5<br><br>**DNS** for L-2 |
| **P6:** x5 function in A7b | A250 | **y, x5** | **Fig A5.** Effects of deleting pair (37, 175) and (151, 175) |
| **A8a, A8b, A9a, A9b** as in A7a, A7b | B250 C250 | y, B(x1,x5, x6) C(x5, x6, x10) | B250: **Fig A6.** IPs in x1<br>C250: **Fig A7.** IPs x10. |
| **P7:** functional forms of variables altered by IPs identified in **A7-A9**; multivariable analysis | 250 (A/B/C) | y, x1, x5, x10 | **Fig 6.** Effects of IPs clearly visible in some variables and in some datasets. |
| **A10a, A10b:** check for IPs in univariable analysis; L-1 and L-2 | A125 | y, x1, x3, x5-x7, x10 | **DNS.** IPs in x1, x6 and x7. |
| **A11a:** check for IPs in multivariable analysis; L-1 | A125 | y, x5, x6 | **DNS.** No IPs found. MFP model included x5, x6 and x8 |
| **A11b:** as in A11a; L-2 | A125 | y, x5, x6 | **DNS.** No IPs found |
| **A12a, A12b:** check for IPs in univariable analysis; L-1 | A500 | y, x1-x10 | **DNS.** (L-1) IPs found in x5, x6 and x7. DNS for (L-2) |
| **A13a, A13b:** check for IPs in multivariable analysis; L-1 and L-2 | A500 | y, x1, x3, x5, x6, x10, x4a, x8, x9b | **DNS.** IPs only in x5. x7 was not selected |
| **A14:** effects of sample size and IPs on selected MFP models | A125 A250 A500 | y, x1-x10 | **Table 4.** MFP models from full data and data without IPs |
| **P8:** functional forms from MFP: Effects of sample size and IPs | A125 A250 A500 | y, x1, x3, x5, x6, x10 | **Fig 7.** x5 and x6 had good agreement with the true function in all data. x1 and x3 not selected in A125. Low power. |
| **A15a, A15b:** as A13a and A13b for other data | B500 | y, x1, x3, x5, x6, x10, x2, x4a, x8 | **DNS.** IPs only in x5. x7 was not selected |
| **A16a, A16b: as A13a** and **A13b** for other data | C500 | y, x1, x3, x5, x6, x10, x4a, x8, x9a | **DNS.** IPs only found in x5. x7 was not selected |
| **A17:** MFP models in full data and in data without IPs identified in **A12**, **A15** and **A16** | 500 (A/B/C) | y, x1, x3, x5, x6, x10, x2, x4a, x8, x9a, x9b | **Table A5.** Good agreement between selected models and the true model. True FP2 for x1 always identified. x3 (true linear) always correct. |
| **P9:** functional forms from MFP | 500 (A/B/C) | y, x1, x3, x5, x6, x10 | **Fig 8.** Functional forms estimated from data without IPs |
| **A18:** function Selection Procedure (FSP); Univariable analysis | Bodyfat | Pcfat, all 13 covariates | **Table A6.** Variables abdomen, weight, ankle and hip with non-linear effect. Linear function chosen for another 8 variables. Variable height is non-significant. |
| **A19a, A19b:** check for IPs in univariable analysis; L-1 and L-2 | Bodyfat | All variables | **DNS.** IPs in ankle, biceps, abdomen, hip and weight. No IPs in other variables. Same results for L-1 and L-2. |
| **A20:** MFP models in full data and data without IPs identified in A19 | Bodyfat | All variables | **Table A7.** No non-linear function after deleting IPs |
| **P10:** functional forms; multivariable analysis of full data | Bodyfat | Pcfat, biceps, abdomen, height, wrist | **Fig A8.** Functional forms for continuous variables selected by MFP (0.05, 0.05). |
| **A21a:** check for IPs in selected MFP (0.05, 0.05) model L-1 | Bodyfat | Pcfat, biceps, abdomen, height,wrist | **DNS.** No IPs found. |
| **A21b:** as 21a, (L-2) | Bodyfat | Pcfat, biceps, abdomen, height and wrist | **DNS.** IPs found in biceps. |

D = description; A = analysis; P-presentation; IPs = Influential Points; L-1 = leave-one-out approach; L-2 = leave-two-out approach; NA = not applicable; DNS = Data not shown

## 3 More Details on Simulated Data

### 3.1 contribution of each variable to the model fit

**Table A2** ART data (N = 5,000, $R^2$=0.49). Contribution of each predictor to the model fit, expressed in terms of the percentage reduction in $R^2$ when regressing the index on all predictors minus the one of interest. The last column shows the variables that was used to generate the outcome variable

| Variable | Model[a] | $R^2(-x_i)$ | % reduction $R^2$ | True model |
|---|---|---|---|---|
| x1 | 0.5, 1 | 0.47 | 4.8 | x |
| x2 | 1 | 0.49 | 0.1 | - |
| x3 | 1 | 0.47 | 5.0 | x |
| x4a | 1 | 0.48 | 3.4 | x |
| X4b | 1 | 0.49 | 0.0 | - |
| x5 | -0.2 | 0.22 | **56.0** | x |
| x6 | 0 | 0.41 | **17.1** | x |
| x7 | 1 | 0.49 | 0.1 | - |
| x8 | 1 | 0.46 | 6.0 | x |
| x9a | 1 | 0.49 | 0.0 | - |
| X9b | 1 | 0.49 | 0.01 | - |
| x10 | 1 | 0.47 | 5.4 | x |

[a] are FP powers, "x" denotes a signal variable while "-" denotes a noise variable

### 3.2 Description of the Data A250

For the data A250, we describe the univariable distribution and the correlation structure in detail. This will provide good insight into the design of the simulated data.

*Univariable Distributions*

In real data, initial data analysis needs to be carried out before the formal statistical analysis (Huebner et al. 2018) in order to check for data errors and whether some manipulations are needed such as combining one or more categories with very low percentages. However, this part is less relevant in simulated data and in all data sets used, we assumed that this work has been carefully done. Table A3 presents descriptive statistics for sample A250. It can be seen that variables $y, x1$ and $x10$ have skewness and kurtosis near 0 and 3 respectively, hence they are approximately normally distributed. Variable x5, x6 and x7 have high kurtosis an indication of existence of extreme values or outliers that warrant investigation. In addition, there are only 16 observations (6%) in category 3 of x9. In building a multivariable model, it may be worth combining it with one of the other two categories, whichever is more sensible from a subject-matter point of view. This would be an issue handled in initial data analysis but we decided against combining any categories and used two dummy variables because x9 is a nominal variable.

**Table A3. Data A250.** Descriptive statistics for continuous (top) and categorical (bottom) variables

| Variable | Mean | Median | SD | Min. | Max. | Skewness | Kurtosis |
|---|---|---|---|---|---|---|---|
| y | 12.17 | 12.10 | 0.99 | 8.28 | 15.15 | -0.04 | 3.72 |
| x1 | 54.24 | 54.00 | 9.69 | 24.00 | 85.00 | 0.16 | 2.98 |
| x3 | 20.78 | 19.00 | 9.31 | 5 | 81.00 | 1.93 | 10.44 |
| x5 | 6.90 | 3.16 | 16.94 | 0.46 | 234.34 | 10.27 | 132.50 |
| x6 | 147.92 | 80.00 | 184.24 | 0.00 | 1200.00 | 2.44 | 10.44 |
| x7 | 105.96 | 54.50 | 169.85 | 0.00 | 1727.00 | 5.26 | 43.36 |
| x10 | 16.95 | 15.90 | 8.32 | 0.61 | 43.46 | 0.70 | 3.35 |

| Variable | 0 | 1 | 2 | 3 |
|---|---|---|---|---|
| x2 | 62 (24.80%) | 188 (75.20%) | | |
| x4 | | 30 (12%) | 164 (65.6%) | 56 (22.4%) |
| x8 | 165 (66%) | 85 (34%) | | |
| x9 | | 176 (70.4%) | 58 (23.2%) | 16 (6.4%) |

*Correlation Structure*

Table A4 provides a summary of Spearman rank correlation coefficients and their differences with Pearson correlation coefficients. Only Spearman correlation coefficients whose absolute values were greater than 0.25 and differences greater than 0.05 were shown. Larger differences between the two types of correlation coefficients may indicate presence of outliers or stronger non-linear relationships. Given that the Spearman correlation is more robust against outliers than the Pearson correlation, the largest difference between y and x5 suggests that variable x5 may contain outliers.

**Table A4**. Data A250. The entries above and below the main diagonal are Spearman correlation coefficients with absolute values larger than 0.25 and differences between Spearman and Pearson correlation coefficients greater than 0.05 for continuous variables.

|     | y     | $x_1$ | $x_3$ | $x_5$ | $x_6$ | $x_7$ | $x_{10}$ |
|-----|-------|-------|-------|-------|-------|-------|----------|
| y   | -     | ·     | ·     | -0.45 | ·     | ·     | ·        |
| $x_1$ | ·   | -     | ·     | ·     | ·     | ·     | ·        |
| $x_3$ | ·   | ·     | -     | -0.30 | ·     | ·     | ·        |
| $x_5$ | -0.25 | ·   | -0.15 | -     | 0.29  | ·     | ·        |
| $x_6$ | ·   | ·     | ·     | 0.10  | -     | 0.40  | ·        |
| $x_7$ | ·   | ·     | ·     | -0.11 | ·     | -     | ·        |
| $x_{10}$ | · | ·    | ·     | ·     | ·     | ·     | -        |

## 4 Results
### 4.1 Univariable Analysis

***Identification of Influential Points in Data Set B250 and C250***

In B250, three IPs (95, 160, and 223) were only identified in variable x7, with obs. 223 having a substantial influence on both the selection of variable x7 and its functional forms. For instance, the removal of this observation made variable x7 be included in the model (Table 2) and the test of FP2 vs. linear significant, implying a non-linear function.

Several IPs were observed in data set C250 (Figure A1). The estimated FP2 (3, 3) function for x1 (Table 2) was purely driven by an influential obs.213, whose removal led to a linear function. Further investigation of six IPs reported in variable x6 using the leave-two-out approach revealed that out of six observations, only obs. 52 was highly influential as expected since the other five observations were borderline significant. Besides obs. 52, other IPs were identified, but the residual plots (Figure A2, lower panel) suggested that a non-linear function was a better fit than a linear function obtained by deleting all IPs; hence only obs. 52 was considered an influential point in variable x6. Furthermore, in C250, observations 133 and 209 were considered IPs in variable x5 in the leave-two-out approach. Removal of these two observations led to an FP2 (0.5, 0.5) function. An inspection of the residual plot (Figure A2, upper panel) supported an FP2 rather than an FP1 (0) function estimated from the complete data, thus observations 133 and 209 were considered IPs. Although several IPs for variable x7 were reported in single case deletion, all of them were inconsequential based on leave-two-out except case 104, which had the largest deviance difference when individually deleted. However, its deletion produced a complex FP2 (2,-0.5) function that was quite similar to the estimated function from the complete data except at the beginning where a hook was visible (Figure A3). Sauerbrei and Royston (1999) experienced a similar outcome and suggested the use of pre-transformation of the variable before fitting an FP model. Here, we retained this observation and allowed the maximum permitted function to be an FP1, which produced an FP1(0) function as shown in Figure A3.

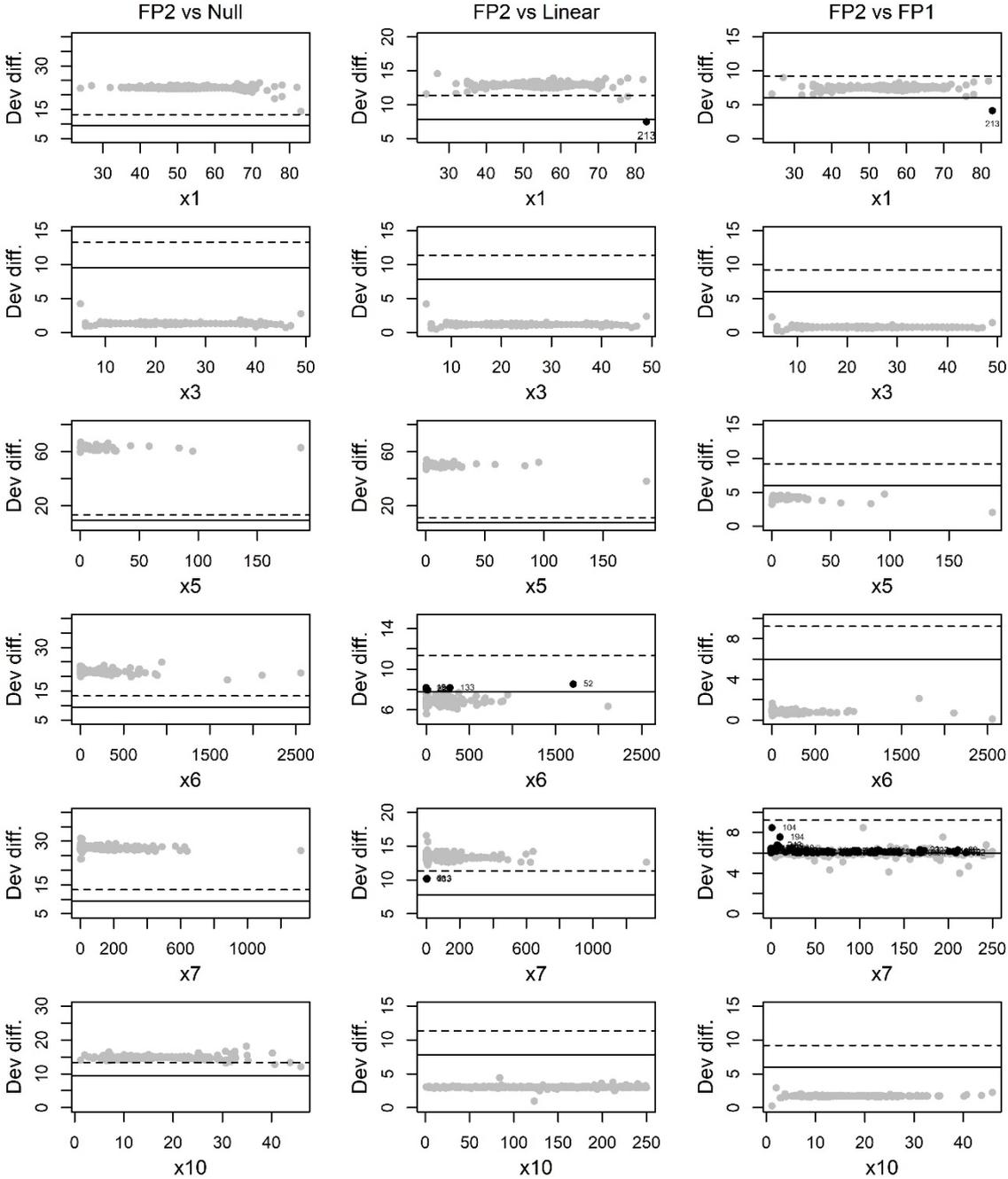

**Figure A1** Data C250. Identification of influential points in univariable analysis using leave-one-out approach

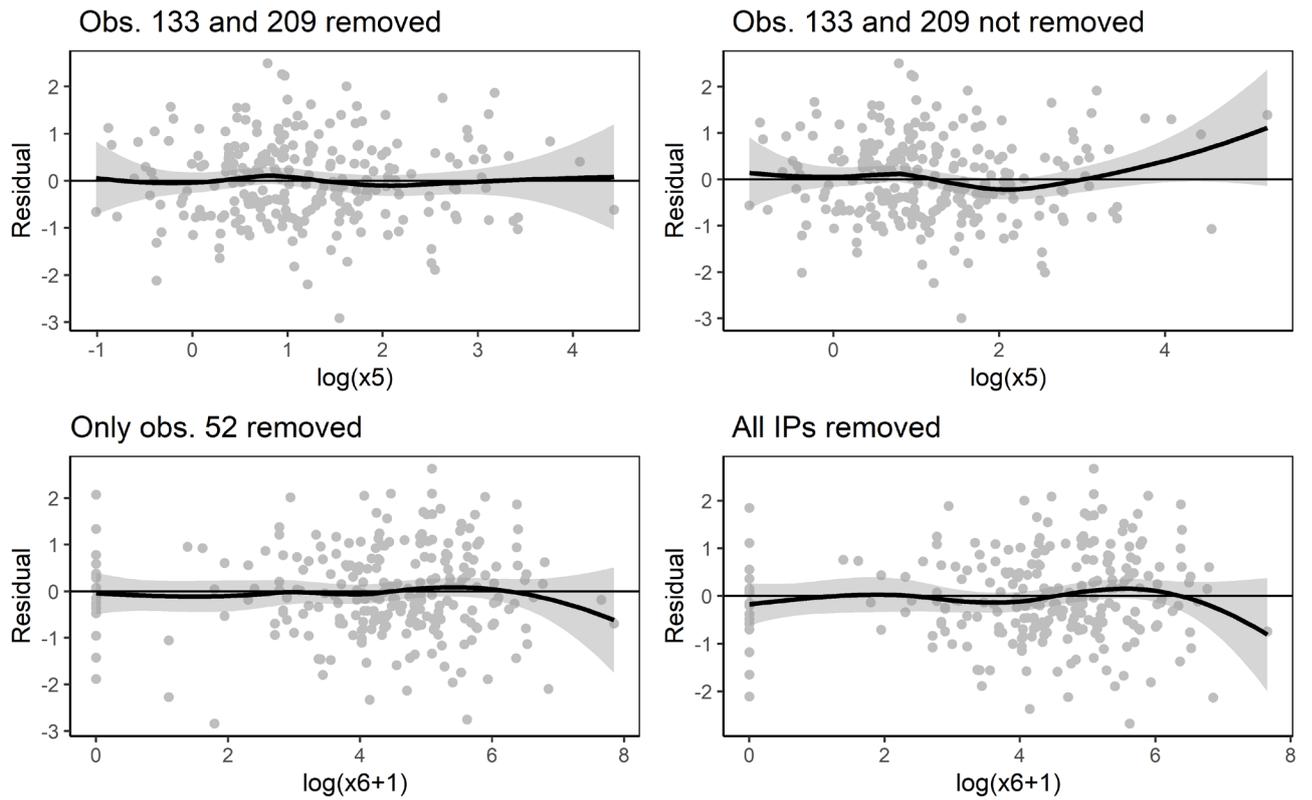

**Figure A2.** Data C250, univariable analysis. Smoothed residuals with 95% pointwise confidence intervals for variable x5 and x6 before and after removal of IPs.

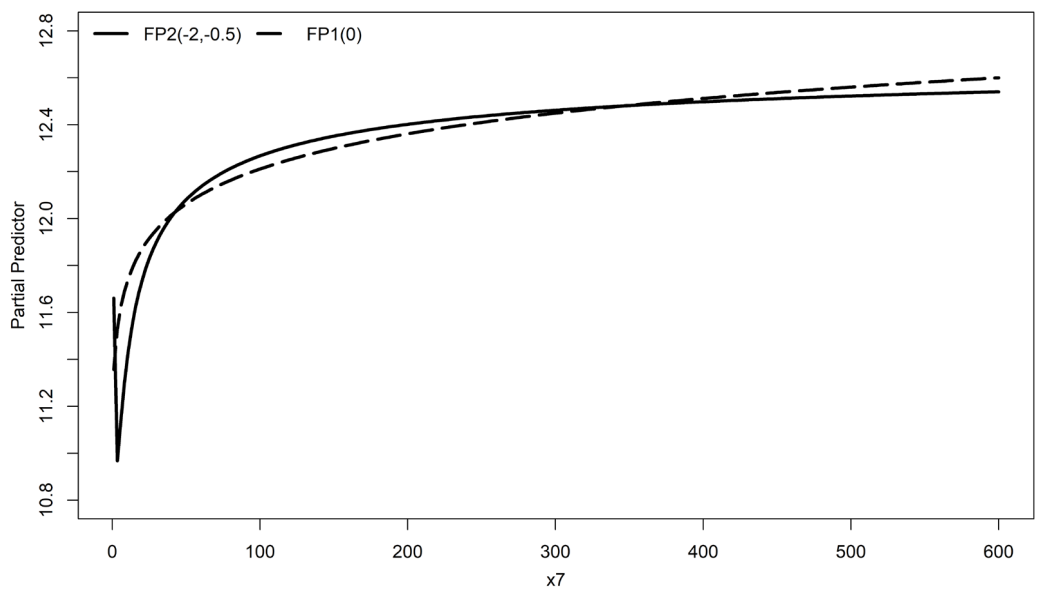

**Figure A3** Data C250. Functional form of variable x7 in full data (dashed line) and without observation 104 (solid line). Truncated at 600.

## 4.2 *Results of Multivariable Analysis*

*Identification of Influential Points in Data A250*

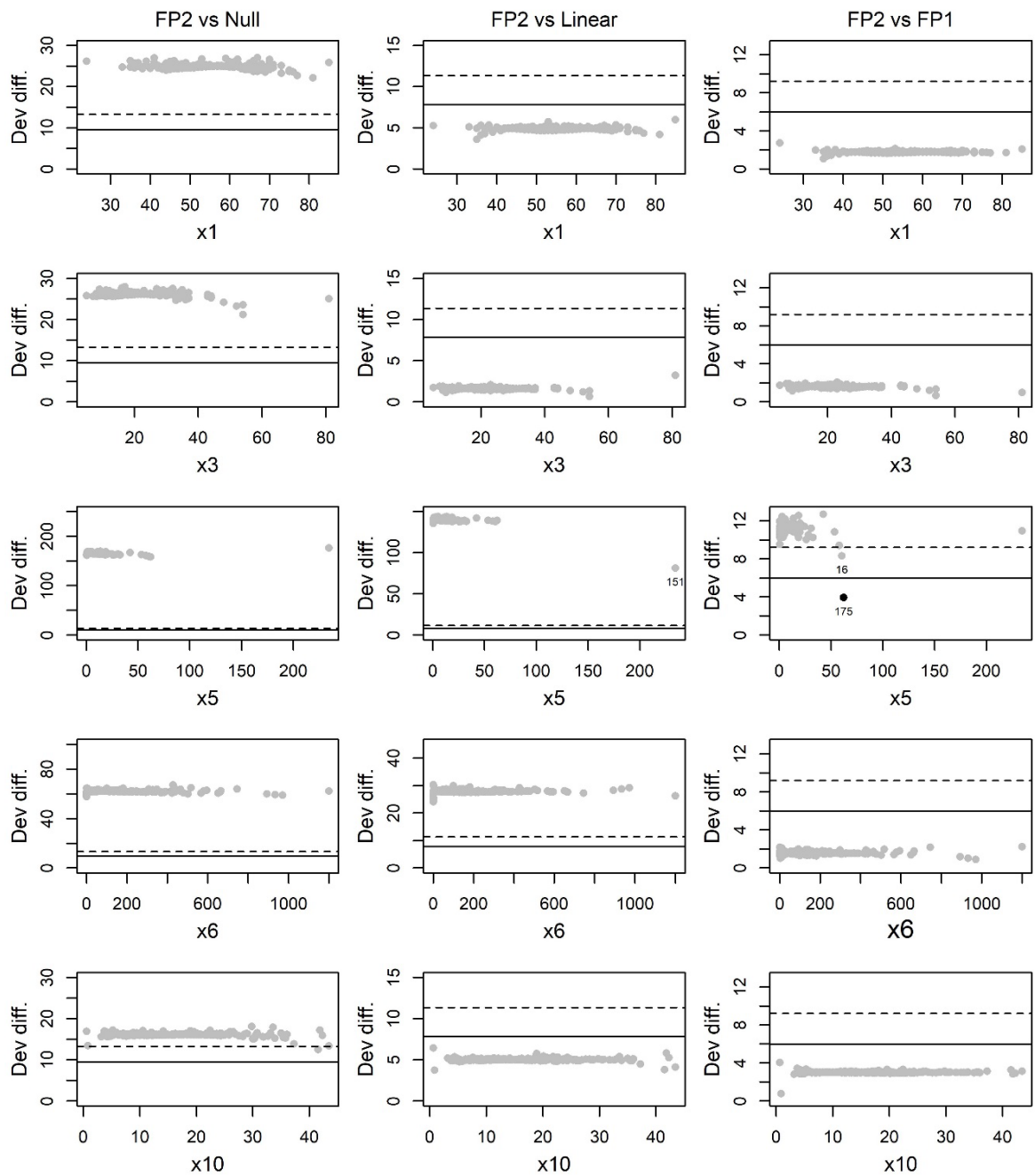

**Figure A4** Data A250. Identification of influential points using L-1 approach in the selected MFP model (see Table 3, all data)

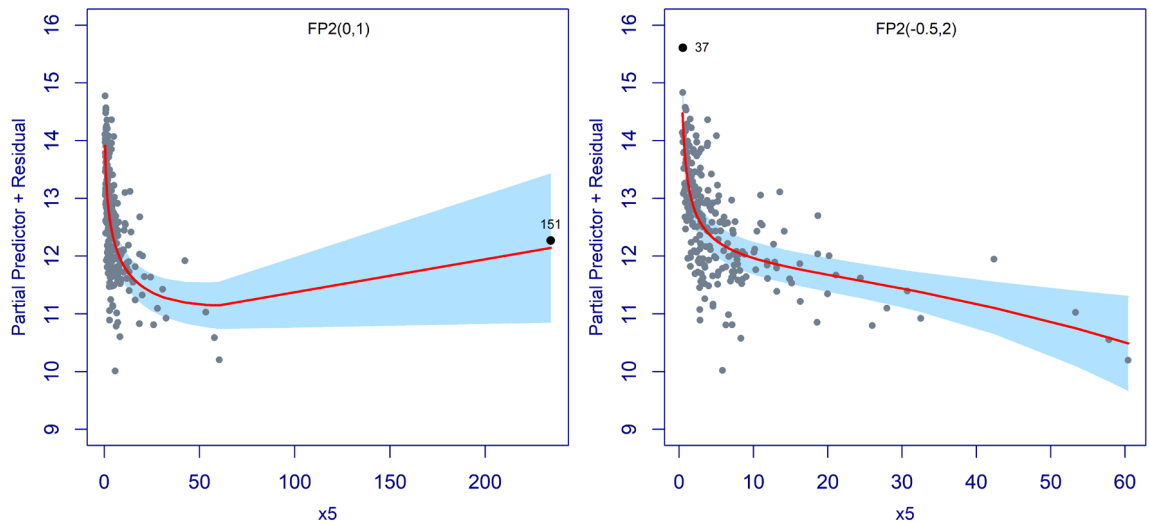

**Figure A5** Data A250. Identification of influential points in multivariable analysis using leave-two-out approach. left panel: functional form for x5 when the pair (37, 175) was removed. Right panel: functional form for x5 when pair (151, 175) was removed.

*Identification of Influential Points in Data B250 and C250*

In data B250, four IPs were solely identified in variable x1 using the leave-one-out approach. The deletion of any of these observations made the test of FP2 vs. FP1 non-significant (Figure A6). Interestingly, the leave-two-out approach disclosed that the test of FP2 vs. FP1 was not necessary since it was only driven by obs. 6. The pair (6, 221) made the test of FP2 vs. linear significant. When these two observations were removed, a linear function was sufficient to describe variable x1. As a result, observations 6 and 221 were considered influential points. Observation 98 was not considered influential, yet it had an extremely large value of variable x5. For this reason, the observation was removed before the functional plots were constructed and similar results were obtained. In total, 3 observations (6, 98, and 221) were completely removed from the analysis, as shown in Table 3. From Table 3, two situations were observed in the comparison of "all", "IPAm" and "true" models. First, a linear function was estimated for variable x1 when the three influential observations were deleted rather than an FP2 (-1, 3) function obtained from complete data. It was obvious that a linear function was not a good approximation of the true FP2 (0.5, 1) function (Figure 6), which can be attributed to the low power of detecting non-linearity. Secondly, when IPs were removed, variable x8 was included in the model, which was initially excluded.

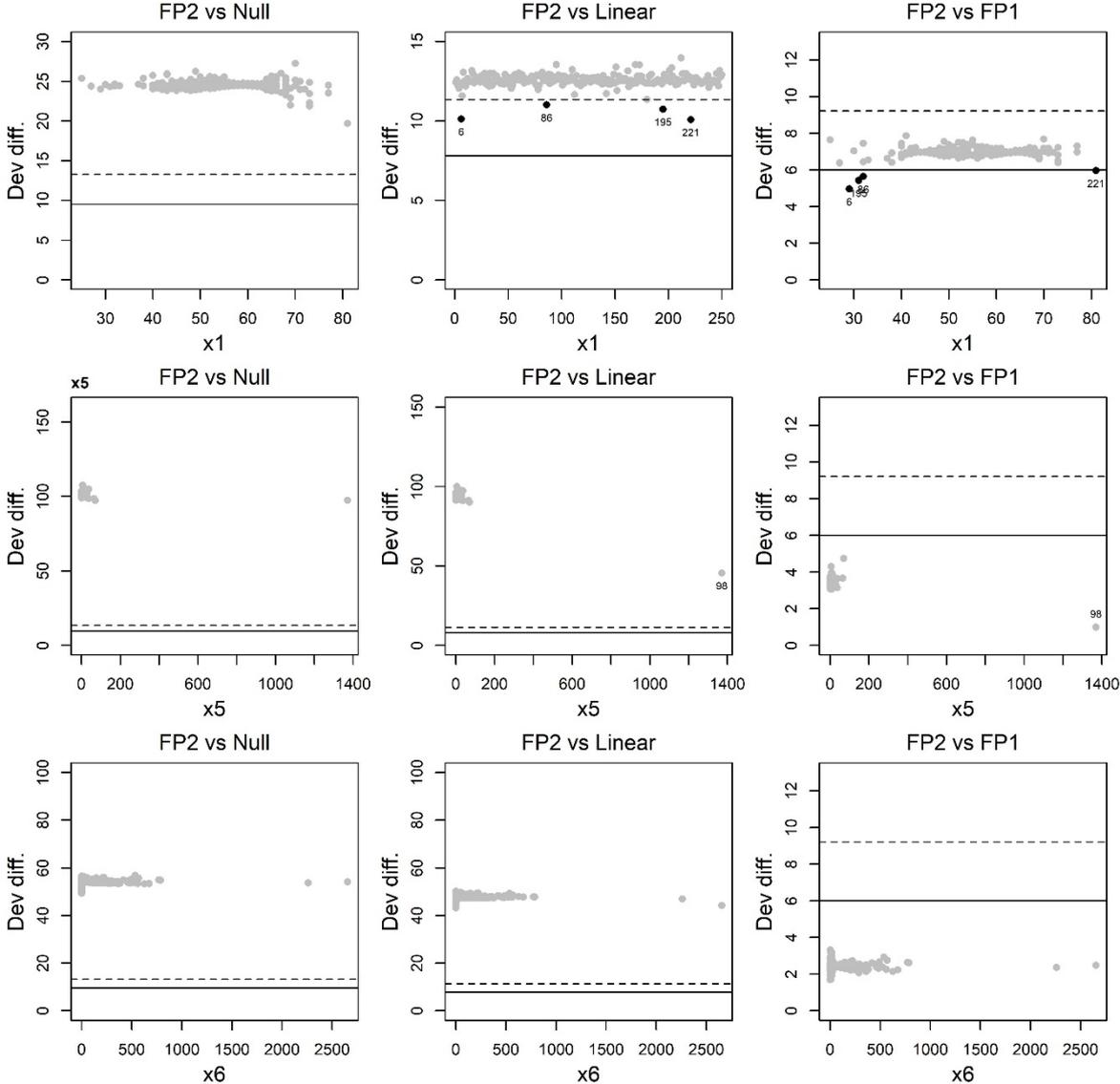

**Figure A6** data B250. Identification of influential points in the selected MFP model (see Table 3). Multivariable analysis using L-1 approach

In data C250, the leave-one-out approach identified several IPs in variable x10 that made the test of FP2 vs. linear non-significant, though most were at borderline except six observations, out of which obs. 138 had the largest influence (Figure A7). Moreover, the leave-two-out approach identified six observations (82, 123, 129, 138, 194, and 214) with substantial influence on variable x10, which also correspond to the first six observations with large deviance differences when individually deleted. Only observation 123 was influential in both univariable and multivariable analysis. From Table 3, the two models ("all" and "IPAm") are similar except that variable x10 was excluded when IPs were deleted.

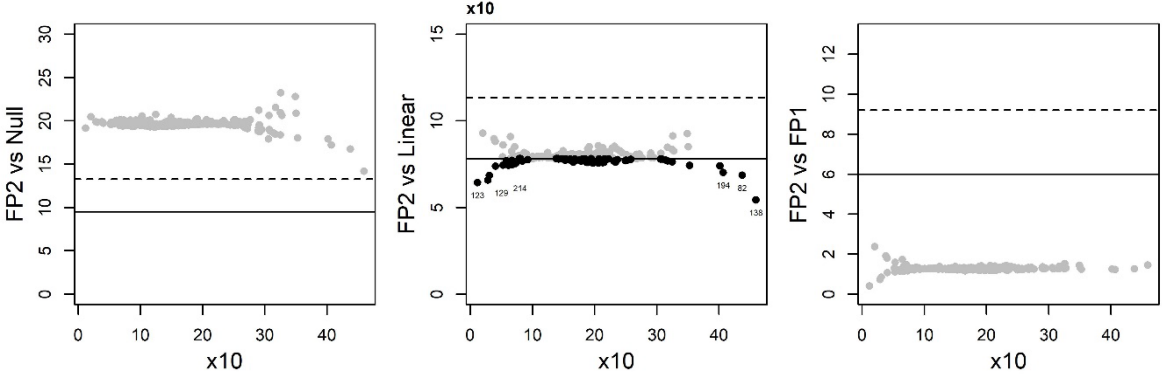

**Figure A7** data C250. Identification of influential points of x10 in the selected MFP model using L-1approach

## 4.3 Identifiability of the MFP models in relatively large datasets

**Table A5** Data A500, B500 and C500. Multivariable analysis for relatively large datasets. See Figure 8 for the related functions.

| Variables | A500(1) All | A500(1) IPAm | B500(1) All | B500(1) IPAm | C500(1) All | C500(1) IPAm | True model |
|---|---|---|---|---|---|---|---|
| x1 | 0.5, 1 | = | -0.5, 3 | 0, 3 | 2, 2 | = | 0.5, 1 |
| x3 | 1 | = | 1 | = | 1 | = | 1 |
| x5 | 0, 3 | 0 | -0.5, -0.5 | 0 | 0, 3 | 0 | -0.2 |
| x6 | 0 | = | 0 | = | 0 | = | 0 |
| x7 | out | = | out | = | out | = | Out |
| x10 | 1 | = | 1 | = | 3 | = | 1 |
| x9a | out | = | in | = | out | = | Out |
| x9b | in | out | out | = | out | = | Out |
| x2 | out | = | out | = | in | = | Out |
| x4a | in | = | in | = | in | = | In |
| x4b | out | in | out | = | out | = | Out |
| x8 | in | = | in | = | in | = | In |

# 5 Case study-body fat data

## 5.1 Univariable Analyses

Table A6 shows the results of the function selection procedure. All variables except height (p=0.994) were considered important predictors of body fat at the 0.05 significance level. Four variables (abdomen, weight, ankle, and hip) were found to be non-linear, as shown in the test of FP2 vs. linear.

Leave-one-out approach identified four IPs (31, 39, 182, and 192), whereas leave-two-out approach identified three IPs (31, 39, and 54). Observations 182 and 192 that were influential in the leave-one-out approach were not influential in the leave-two-out approach probably because they were borderline significant. Removal of obs. 39 or any pair with this observation rendered the test of FP2 vs. linear non-significant in univariable models of the abdomen, hips, and weight. Removal of obs.31 in the ankle rendered the test of FP2 vs. linear non-significant. This implies that in univariable analysis, the linearity assumption sufficed for these variables.

Interesting results were found in a univariable model of biceps; deleting any pair with obs.39 produced a non-linear function except when 11 pairs were deleted. Out of these 11 pairs, a pair (39, 54) decreased the deviance difference considerably. Therefore, instead of deleting all the observations that comprise 11 pairs, which perhaps might lead to a loss of power, only pair (39, 54) was deleted. Hence, observations 39 and 54 were considered IPs, their deletions produced a linear function for the biceps (Table A6). Once all IPs were deleted (obs. 31, 39, and 54), it turned out that no covariate was non-linearly related to the outcome variable, as can be seen in Table A6 (last column).

Table A6: Data body fat, univariable analysis. P-values for different model comparison are displayed in column 2-4. The last two columns show the FP powers or exclusion of a variable in the complete data set and after deleting influential points respectively.

| Variable | FP2 vs. Null | FP2 vs. Linear | FP2 vs. FP1 | Univariable selection | Univariable selection after 3 IPs deleted* |
|---|---|---|---|---|---|
| Biceps | 0.000 | 0.229 | 0.903 | 1 | 1 (39, 54) |
| Abdomen | 0.000 | **0.000** | 0.036 | **3,3** | 1(39) |
| Height | **0.994** | 0.995 | 0.967 | Out | Out |
| Wrist | 0.000 | 0.862 | 0.906 | 1 | 1 |
| Weight | 0.000 | **0.019** | 0.786 | **-0.5** | 1(39) |
| Neck | 0.000 | 0.811 | 0.644 | 1 | 1 |
| Thigh | 0.000 | 0.079 | 0.932 | 1 | 1 |
| Ankle | 0.000 | **0.044** | 0.165 | **-2** | 1(31) |
| Forearm | 0.000 | 0.390 | 0.228 | 1 | 1 |
| Age | 0.000 | 0.808 | 0.673 | 1 | 1 |
| Chest | 0.00 | 0.518 | 0.819 | 1 | 1 |
| Hip | 0.000 | **0.008** | 0.575 | **-2** | 1(39) |
| Knee | 0.000 | 0.864 | 0.971 | 1 | 1 |

*The number enclosed in brackets are the identified influential points

## 5.2 Multivariable Analyses

The MFP (0.05, 0.05) was applied to complete body fat data, and four variables were selected: biceps, abdomen, height, and wrist. These variables explained about 75% of the total variation as shown in Table A7. Analysis of the contribution of each variable to the model fit showed that the abdomen was the most important predictor since its omission led to a reduction of $R^2$ by about 60%. Similarly, its contribution to the "full" model obtained with MFP (1, 0.05) was also large (about 17%). Biceps was the only non-linear variable with an FP2 (3, 3) function. An inspection of its functional form (Figure A8) showed that an obs. 39 with the largest value of biceps caused an FP2 function. Hence, there is a need to check for influential points.

In the univariable approach, three IPs (31, 39, and 54) were identified. After removing them, a different MFP model was selected (see Table A7, column IPBFu) where biceps and height were excluded, while weight and thigh were included with linear and FP2 (-2, -2) functions, respectively.

In this subsection, we present the results of the diagnostic check of IPs in the selected MFP model with four variables: biceps (3,3), abdomen (1), height (1), and wrist (1). The leave-one-out approach did not identify any influential points, while the leave-two-out approach identified IPs in the biceps. Deleting any pair with observation 39 made the test of FP2 vs. null borderline significant except that it became non-significant when 17 pairs were deleted. Moreover, deleting two pairs (39, 172) and (39, 180) rendered the test of FP2 vs. linear non-significant. This implies that, besides obs.39, other influential points existed in the data. Among the 17 pairs, deleting a pair (39, 172) decreased the deviance difference the most (in FP2 vs. null), followed by (39, 180), while others were close to the borderline. To avoid loss of information resulting from deleting all 17 pairs, we only considered observations 39, 172, and 180 as influential points in the multivariable model. The results are summarized in Table A7 (column labelled IPBFm). The removal of these three IPs led to the exclusion of biceps and height, while weight was included in the model. Furthermore, all the selected continuous variables were described by a linear function, so non-linearity was driven by IPs.

**Table A7** Data body fat. Selected models from a MFP analysis with all data represented by MFP(0.05, 0.05) and after removal of IPs identified from univariable (IPBFu (k)) and multivariable (IPBFm(l)) diagnostic analyses where k and l represent the number of IPs identified in the corresponding analysis. MFP (1, 0.05) – no elimination of variables, FSP with significance level 0.05.

|  | MFP (0.05, 0.05) | | MFP (1, 0.05) | | IPBFu (3)* | IPBFm (3)* |
|---|---|---|---|---|---|---|
| Variable | Model[a] | $R^2_{red}$ | Model[a] | $R^2_{Red}$ | Model[a] | Model[a] |
| Biceps | 3, 3 | 4.32 | 1 | 0.09 | out | = |
| Abdomen | 1 | 59.82 | 0 | 16.79 | 1 | = |
| Height | 1 | 1.69 | 1 | 0.37 | out | = |
| Wrist | 1 | 2.63 | 1 | 1.52 | 1 | = |
| Weight | Out | - | 1 | 0.01 | 1 | = |
| Neck | Out | - | 1 | 0.63 | out | = |
| Thigh | Out | - | 1 | 0.12 | -2, -2 | out |
| Ankle | Out | - | 1 | 0.07 | out | = |
| Forearm | Out | - | 1 | 0.41 | out | = |
| Age | Out | - | 1 | 0.54 | out | = |
| Chest | Out | - | 1 | 0.08 | out | = |
| Hip | Out | - | 1 | 0.12 | out | = |
| Knee | Out | - | 1 | 0.01 | out | = |
| $R^2$ | 0.748 | | 0.757 | | 0.745 | 0.737 |

[a] Numbers are FP powers; [b] $R^2_{Red}$ denotes percentage reduction in $R^2$ by eliminating a variable from the selected MFP model (see also Table A2). IPs 31, 39 and 54 were identified in IPBFu while obs.39, 172 and 180 were identified in IPBFm.

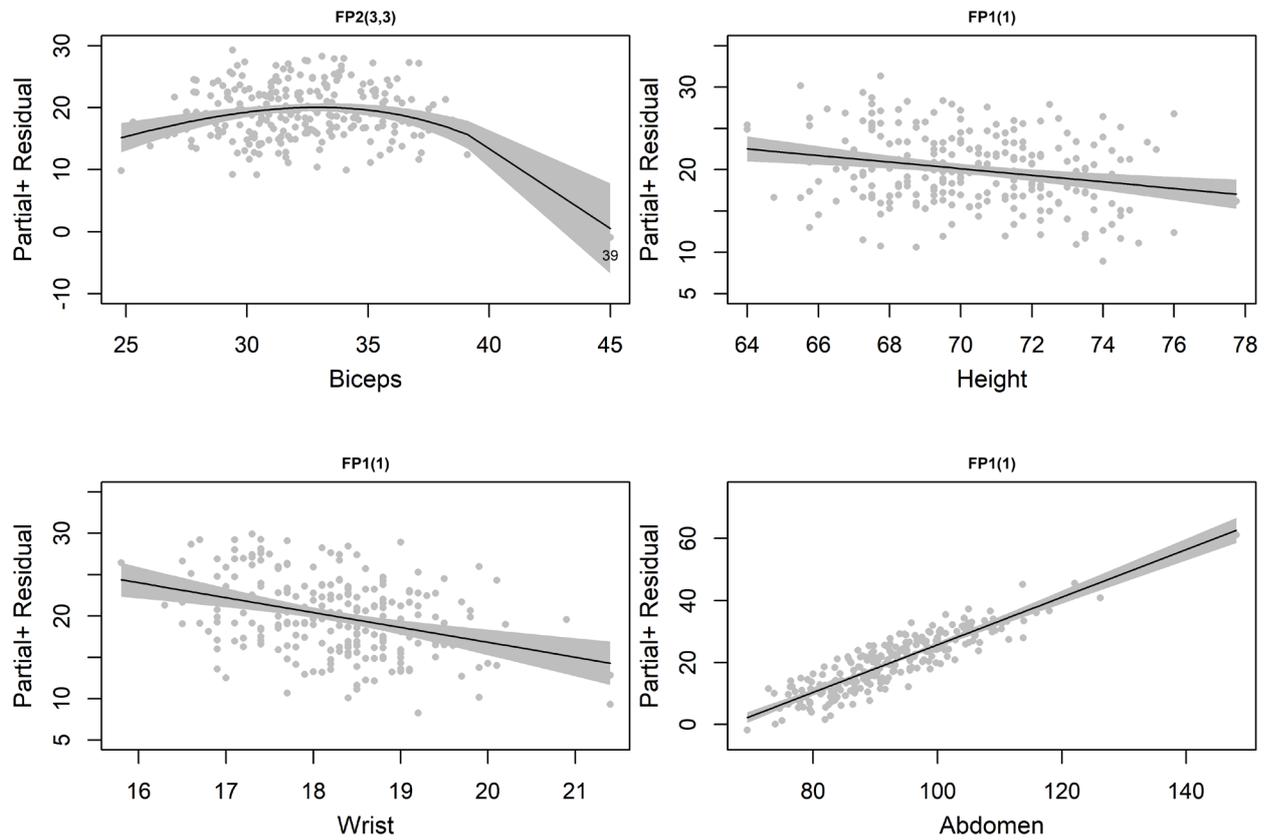

**Figure A8** Data body fat. Multivariable analysis of complete data. Functional forms for continuous predictors in MFP (0.05, 0.05) model. Deleting 3 IPs biceps is no longer significant.